\documentclass[a4paper,11pt]{article}

\usepackage{jheppub} 

\usepackage{graphicx}

\usepackage{amsmath}
\usepackage[utf8]{inputenc}

\usepackage{amssymb}
\usepackage{amsfonts}
\usepackage{amsbsy}
\usepackage{url}
\usepackage{enumerate}
\usepackage{caption}
\usepackage{color}
\usepackage{ulem}

\newcommand{\be}{\begin{eqnarray}}
\newcommand{\ee}{\end{eqnarray}}
\newcommand{\p}{\partial}

\makeatletter
\@addtoreset{equation}{section}
\makeatother

\newcommand\rsout{\bgroup\markoverwith{\textcolor{red}{\rule[0.5ex]{2pt}{0.4pt}}}\ULon}

\title{Non-Abelian strings and domain walls in 
two Higgs doublet models}

\author[a]{Minoru Eto,}
\author[b,c]{Masafumi Kurachi}
\author[d,b]{and Muneto Nitta}

\affiliation[a]{Department of Physics, Yamagata University,\\
Kojirakawa-machi 1-4-12, Yamagata, Yamagata 990-8560, Japan}
\affiliation[b]{Research and Education Center for Natural Sciences, Keio University,\\
Hiyoshi 4-1-1, Yokohama, Kanagawa 223-8521, Japan}
\affiliation[c]{Theory Center, High Energy Accelerator Research Organization (KEK),\\ Oho 1-1, Tsukuba, Ibaraki 305-0801, Japan}
\affiliation[d]{Department of Physics, Keio University,\\
Hiyoshi 4-1-1, Yokohama, Kanagawa 223-8521, Japan}

\emailAdd{meto(at)sci.kj.yamagata-u.ac.jp}
\emailAdd{kurachi(at)keio.jp}
\emailAdd{nitta(at)phys-h.keio.ac.jp}

\abstract{
Contrary to the standard model that does not admit topologically nontrivial solitons, two Higgs doublet models admit
topologically stable vortex strings and domain walls. 
We numerically confirm the existence of a topological $Z$-string confining fractional $Z$-flux inside. We show that topological strings at $\sin\theta_W = 0$ limit reduce to non-Abelian strings which possess non-Abelian moduli $S^2$ associated with spontaneous breakdown of the $SU(2)$ custodial symmetry. We numerically solve the equations of motion for various parameter choices. It is found that a gauging $U(1)_Y$ always lowers the tension of the $Z$-string while it keeps that of the $W$-string.
 On the other hand, a deformation of the Higgs potential is either raising or lowering the tensions of the $Z$-string and $W$-string. We numerically obtain an effective potential for the non-Abelian moduli $S^2$ for various parameter deformations under the restriction $\tan\beta=1$. It is the first time to show that there exists a certain parameter region where the topological $W$-string can be the most stable topological excitation, 
contrary to conventional wisdom of electroweak theories.
We also obtain numerical solutions of composites of the string and domain walls in a certain condition.
}
\preprint{YGHP-18-08, KEK-TH-2054}

\begin{document}
\maketitle


\section{Introduction}

The discovery of the Higgs boson at the Large Hadron Collider (LHC) at CERN~\cite{Aad:2012tfa, Chatrchyan:2012xdj} and subsequent measurements on properties of the Higgs boson proved that the Standard Model (SM) is currently the best description of the physics of elementary particles. Still, there is a need to look for physics beyond the SM to solve problems that are left unanswered by the SM, such as masses of neutrinos, baryon asymmetry of the Universe, the origin of the dark matter, etc. Among various possibilities, the Two Higgs doublet model (2HDM)~\cite{Branco:2011iw} is a popular extension of the SM. It introduces two Higgs doublet fields ($\Phi_1$, $\Phi_2$), instead of one as in the case of the SM. Though it is a simple extension of the Higgs sector of the SM, phenomenology of the model is quite rich thanks to the existence of  four additional scalar degree of freedom (charged Higgs bosons ($H^\pm$), CP-even Higgs bosons ($H$) and CP-odd Higgs boson ($A$)), which can be in principle produced at the LHC. 
(See, e.g., Refs.~\cite{Kanemura:2014bqa,Bernon:2015qea,Kling:2016opi,globalfit} and references therein for phenomenological studies of 2HDM.)
Two Higgs doublet fields are also required when one considers supersymmetric extension of the SM \cite{Nilles:1983ge}. 

One of interesting features of the 2HDM is a nontrivial topology of the order parameter space allowing the existence of various topological objects, 
which would indicate important cosmological consequences, 
in contrast to the SM which does not allow topological solitons 
as summarized below. 
A pioneer work on solitons in the SM was done by Nambu \cite{Nambu:1977ag}, in which a classical configuration
of a pair of magnetic monopoles bounded by a $Z$-flux tube was found.
Unfortunately, the configuration 
cannot be stable reflecting the fact that the SM is topologically trivial. 
After a while, a new type of soliton, so-called semi-local strings, has been found in the extended Abelian-Higgs model 
which corresponds to the $\theta_W =0$ limit of the SM   
\cite{Vachaspati:1991dz}. It is peculiar that a semi-local string is stable 
in certain parameter region of topologically trivial models \cite{Vachaspati:1991dz,Hindmarsh:1991jq,Achucarro:1992hs}.
It soon lead to a search of a stable non-topological soliton in the SM \cite{Vachaspati:1992fi,Vachaspati:1992jk}, and 
the so-called electro-weak(EW) $Z$- and $W$-string were found \cite{Vachaspati:1992jk}. 
However the EW strings are stable only in a parameter region with $\sin^2\theta_W \lesssim 1$ which is quite 
far from the realistic point $\sin^2\theta_W \simeq 0.23$ \cite{Vachaspati:1992jk,James:1992zp,James:1992wb}, however
various interesting aspects of the EW strings can be found in Ref.~\cite{Achucarro:1999it}. 
For instance, 
$Z$-strings were suggested to contribute to electroweak baryogenesis
\cite{Brandenberger:1992ys,Barriola:1994ez} (see also Ref.~\cite{Nagasawa:1996gy}), and
$Z$-strings ending on monopoles were suggested to generate 
primordial magnetic fields in cosmology 
  \cite{Vachaspati:2001nb,Poltis:2010yu}.
Monopole and anti-monopole connected by a $Z$-string 
 can be saddle point solutions, known as sphalerons \cite{Klinkhamer:1984di}.

Let us turn back to the 2HDM.
EW strings in the 2HDM were studied around the almost same period \cite{La:1993je,Earnshaw:1993yu,Perivolaropoulos:1993gg,Bimonte:1994qh}. 
(See also \cite{Ivanov:2007de} for more recent study.)
A membrane \cite{Bachas:1995ip,Riotto:1997dk} and
sphaleron(-like soliton) \cite{Bachas:1996ap,Grant:1998ci,Grant:2001at,Brihaye:2004tz} were also studied.
Topological defects such as domain walls, global vortices, and global monopoles were studied in Refs.~\cite{Battye:2011jj, Thesis}. (See also Ref.~\cite{Bachas:1998bf}.)
Among them, EW strings are quite similar to those in the SM in the sense that they are not topologically protected. 
In contrast, it was first pointed out in Ref.~\cite{Dvali:1993sg} that the 2HDM admits a topologically stable EW string solution 
associated with the spontaneously broken 
$U(1)_a$ symmetry, 
which is the difference between the overall phases of the two Higgs fields. 
The topological EW string has two distinguishable aspects: 
It is a global string whose winding partially goes inside 
the global phase $U(1)_a$,
and it is, at the same time, a local string in the sense that the rest of winding is supplemented by an $SU(2)_{W}$ gauge transformation, 
thereby leading a fractional magnetic flux of a $Z$-boson confined in the vortex core \cite{Dvali:1993sg}.
Such strings are also present in supersymmetric extension of SM \cite{Dvali:1994qf}.
This is the unique topologically stable string-like objects in the 2HDM 
playing a role of cosmic strings, 
and therefore it should be an important piece to characterize  vast
parameter space of the 2HDM. Nevertheless, after its discovery, study on the topological $Z$-string has been in a large dormant period
until very recently.
In Ref.~\cite{Eto:2018hhg}, we showed that in a certain parameter space of the 2HDM where the $U(1)_a$ is explicitly broken, 
stable domain walls must appear and be attached to the topologically stable $Z$-strings. 

In this paper, we study theoretical properties of vortex strings and domain walls in the 2HDM in detail.
While Ref.~\cite{Dvali:1993sg} pointed out the existence of the topologically stable string configurations, neither analytic nor numerical solutions were given.
One of the aims of this paper is to confirm the existence of the topologically stable string solutions by solving numerically equations of motion of the 2HDM. 
With the numerical solutions at hand, we will reveal various properties of the strings. 
Firstly, we will point out that the topological $Z$-string is 
a so-called non-Abelian string 
at the limit of $\sin^2\theta_W = 0$ in the
parameter region where the Higgs potential has an exact custodial symmetry.
Non-Abelian strings were extensively studied in supersymmetric gauge theories 
\cite{Hanany:2003hp, Auzzi:2003fs, Auzzi:2003em, Hanany:2004ea, Shifman:2004dr, Gorsky:2004ad, Eto:2005yh, Eto:2006cx, Eto:2006db} 
(see Refs.~\cite{Tong:2005un,Eto:2006pg,Shifman:2007ce} as a review), 
color-flavor locked phase in dense QCD \cite{Balachandran:2005ev,Nakano:2007dr,Nakano:2008dc, Eto:2009kg,Eto:2009bh,Eto:2009tr, Hirono:2010gq, Yasui:2010yw,Fujiwara:2011za,Eto:2011mk,Vinci:2012mc, Cipriani:2012hr, Kobayashi:2013axa,Chatterjee:2015lbf,Alford:2016dco,Chatterjee:2016ykq,Chatterjee:2016tml,Alford:2018mqj}  (see Ref.~\cite{Eto:2013hoa} as a review),
and recently in  the Georgi-Machacek model \cite{Chatterjee:2018znk}. 
As the same with these cases, 
a non-Abelian string in 2HDM has infinitely degenerate family, 
characterized by a moduli pace isomorphic to $S^2 \simeq SU(2)/U(1)$.
This is parameterized by Nambu-Goldstone (NG) modes localized around the non-Abelian string, 
as a consequence of spontaneous breakdown of the custodial symmetry in the presence of the non-Abelian string. 
Points of the moduli space correspond to a magnetic flux of the $SU(2)_{W}$ gauge field, and there are two $Z$-strings corresponding to the north and south poles
of the moduli space $S^2$ while the $W$-strings correspond to the equator of $S^2$. 
They are physically the same solutions which
are transformed to each other by the custodial symmetry.
Once we switch on the $U(1)_Y$ gauge coupling ($\sin^2\theta_W \neq 0$), 
these strings are distinguished. 
First let us consider a Higgs potential 
which is exactly symmetric under the custodial symmetry.
The $U(1)_Y$ is a subgroup of the custodial symmetry, and consequently the Lagrangian no longer has the custodial symmetry.
As a consequence, NG modes localized around the string acquire a mass to become pseudo-NG modes, and
almost all the non-Abelian strings have larger tension  than the two $Z$-strings corresponding to the north and south poles of the $S^2$ moduli space.
Especially, we will make an ansatz for unstable configurations corresponding to generic points of $S^2$, and will be succeed in obtaining 
an effective potential on the $S^2$ moduli. It will turned out that the gauging $U(1)_Y$ lowers tension of the strings. The $Z$-strings
are lowered most while the $W$-strings are intact. Therefore, the $Z$-strings are the most stable string solutions.
Next, we will modify the Higgs potential under the restriction with $\tan\beta = 1$ being kept, in which case the vacuum has the custodial symmetry although the potential does not.
Interestingly, it will turn out that a certain modification of the Higgs potential works opposite to the $U(1)_Y$ gauging. Namely,
it can lift up the tension of $Z$-strings while that of $W$-strings are lowered. 
In a generic point in the parameter space, the two opposite effects compete. We will numerically solve the equations of motion and
will show that there exists a certain parameter region where the $W$-strings are the most stable strings, 
in contrast to the previous work \cite{Dvali:1993sg} considering only the $Z$-strings.
We will proceed our numerical analysis for the most generic Higgs potential for $\tan\beta \neq 1$. We will obtain numerical solutions 
for the $Z$-strings, which is a numerical confirmation of the findings in Ref.~\cite{Dvali:1993sg}.
Finally, we will introduce additional interactions which explicitly break $U(1)_a$,
by giving rise to a double sine-Gordon potential on it.
It gives rise to domain walls ending on the $Z$-strings 
  \cite{Kibble:1976sj, Kibble:1982dd, Vilenkin:1982ks, Everett:1982nm, Vilenkin:2000jqa}, 
as the case of axion strings \cite{Kawasaki:2013ae} 
and axial strings in dense QCD \cite{Eto:2013bxa}.
We will classify the parameter space into three regions;
in which a $Z$-string is attached by one domain wall,  
two domain walls with different tension, 
and one composite domain wall made of two constituent domain walls.
We will make numerical configurations for these three typical string-wall composites.

This paper is organized as follows.
Sec.~\ref{sec:model} is devoted for reviewing the 2HDM with special emphasis on the custodial symmetry which
is peculiar to the 2HDM.
We will investigate the non-Abelian strings at $\sin^2\theta_W = 0$ in Sec.~\ref{sec:NAstring}.
The effects of the $U(1)_Y$ gauging is studied in Sec.~\ref{sec:U(1)Y}, 
and the modification of the Higgs potential (under the
constraint of $\tan\beta=1$) is investigated in Sec.~\ref{sec:mod-Higgs}. 
The $Z$-string solutions for the most generic Higgs potential
is obtained in Sec.~\ref{sec:tanbeta}. The string-wall composites are given in Sec.~\ref{sec:string-wall}. Sec.~\ref{sec:summary} is devoted for summary 
and discussion.


\section{Two Higgs doublet models}\label{sec:model}

We introduce two $SU(2)$ doublet Higgs fields, $\Phi_1$ and $\Phi_2$, with the hypercharge $Y=1$, and consider the following Lagrangian: 
\be
{\cal L} &=& - \frac{1}{4}B_{\mu\nu}B^{\mu\nu} - \frac{1}{4}W_{\mu\nu}^aW^{a\mu\nu} + \sum_{i=1,2}
\left(D_\mu \Phi_i^\dagger D^\mu\Phi_i\right) - V.
\ee
Here, $B_{\mu\nu}$ and $W_{\mu\nu}^a$ represent the field strength of the hypercharge and the weak gauge fields, and $D_\mu$ represents the covariant derivative acting on two Higgs doublet fields. $V$ is the Higgs potential which has the following form:
\be
V &=& 
m_{11}^2\Phi_1^\dagger\Phi_1 
+ m_{22}^2 \Phi_2^\dagger \Phi_2 
- \left(m_{12}^2 \Phi_1^\dagger \Phi_2 + {\rm h.c.}\right) 
+ \frac{\beta_1}{2}\left(\Phi_1^\dagger\Phi_1\right)^2
+ \frac{\beta_2}{2}\left(\Phi_2^\dagger\Phi_2\right)^2 \nonumber\\
&& + \beta_3\left(\Phi_1^\dagger\Phi_1\right)\left(\Phi_2^\dagger\Phi_2\right) 
+ \beta_4 \left(\Phi_1^\dagger\Phi_2\right)\left(\Phi_2^\dagger\Phi_1\right)
+\left\{\frac{\beta_5}{2}\left(\Phi_1^\dagger\Phi_2\right)^2 + {\rm h.c.}\right\},
\label{eq:pot}
\ee
Here, in order to avoid tree-level Higgs-mediated  FCNCs, we imposed a softly-broken $Z_2$ symmetry, $\Phi_1 \to +\Phi_1$ and $\Phi_2 \to -\Phi_2$, on the potential.
Without loss of generality,
$m_{12}$ is taken to be real by rephasing the scalar fields. 
We also assume $\beta_5$ to be real, so that  the  Higgs sector is explicitly CP conserving.
We further assume that the Higgs fields develop non-zero vacuum expectation values (VEVs) as
\be
\Phi_1 = \left(
\begin{array}{c}
\phi_{1,1}\\
\phi_{1,2}
\end{array}
\right) =  \left(\begin{array}{c} 0 \\ v_1\end{array}\right),\quad
\Phi_2 = \left(
\begin{array}{c}
\phi_{2,1}\\
\phi_{2,2}
\end{array}
\right) =  \left(\begin{array}{c} 0 \\ v_2\end{array}\right).
\ee
This occurs when at least one eigenvalue of the mass matrix is negative, then we can set $v_1$ and $v_2$
are simultaneously positive without loss of generality.

For later use, we introduce the following two-by-two matrix form~\cite{Grzadkowski:2010dj} which consists of two Higgs doublet fields:
\be
H = \left( i\sigma_2 \Phi_1^*,\ \Phi_2\right) = \left(
\begin{array}{cc}
\phi_{1,2}^* & \phi_{2,1}\\
-\phi_{1,1}^* & \phi_{2,2}
\end{array}
\right).
\ee
The electroweak gauge transformation acting on this matrix field can be expressed as 
\be
H  &\to& \exp\left(\frac{i}{2}\alpha_a(x) \sigma_a\right) H \exp\left(-\frac{i}{2}\beta(x)\sigma_3\right),
\ee
where $SU(2)_W$ and $U(1)_Y$ gauge transformations act on $H$ from left and right, respectively. The covariant derivative acting on $H$ is thus expressed as
\be
D_\mu H &=& \p_\mu H - g\frac{i}{2}\sigma_a W_\mu^a H + g'\frac{i}{2}H\sigma_3B_\mu.
\ee
The custodial transformation $SU(2)_{C}$ in the two Higgs doublet model \cite{Grzadkowski:2010dj,Pomarol:1993mu} is
identified with a global $SU(2)_{R}$ transformation acting on $H$ from the right 
together with the $SU(2)_W$ transformation as
\be
H \to U^\dagger H U,\quad U\in SU(2)_{C}.
\ee
The potential $V$ in Eq.~(\ref{eq:pot}) can be rewritten as:
\be
V
&=& \frac{m_{11}^2 + m_{22}^2}{2} {\rm Tr}(H^\dagger H )
- \frac{m_{11}^2 - m_{22}^2}{2} {\rm Tr}\left(H^\dagger H \sigma_3\right)
-  m_{12}^2 \left( \det H + {\rm h.c.}\right)\nonumber\\
&+& \frac{2(\beta_1+\beta_2)+3\beta_3}{12}{\rm Tr}\left(H^\dagger H H^\dagger H\right) 
+ \frac{2(\beta_1+\beta_2)-3\beta_3}{12}{\rm Tr}\left(H^\dagger H\sigma_3 H^\dagger H\sigma_3\right)\nonumber\\
&-& \frac{\beta_1 - \beta_2}{3} {\rm Tr}\left(H^\dagger H\sigma_3 H^\dagger H\right) 
+ \left(\beta_3 + \beta_4\right) \det (H^\dagger H)
+ \left(\frac{\beta_5}{2}\det H^2 + {\rm h.c.}\right),
\label{eq:pot_H}
\ee
by using following relations between $H$ and $\Phi$ fields:
\be
{\rm Tr} \left( H^\dagger H \right) &=& \Phi_1^\dagger \Phi_1 + \Phi_2^\dagger \Phi_2,\\
{\rm Tr} \left(H^\dagger H \sigma_3\right) &=& -\Phi_1^\dagger \Phi_1 + \Phi_2^\dagger \Phi_2,\\
\det H &=& \Phi_2^\dagger \Phi_1,\\
{\rm Tr}\left[\left(H^\dagger H\right) \left(H^\dagger H\right)\right]  &=& \left(\Phi_1^\dagger\Phi_1 + \Phi_2^\dagger\Phi_2\right)^2 -2
\left(\Phi_1^\dagger\Phi_2\right) \left(\Phi_2^\dagger\Phi_1\right),\\
{\rm Tr}\left[\left(H^\dagger H\right)\sigma_3 \left(H^\dagger H\right)\right]  &=&
-\left(\Phi_1^\dagger\Phi_1\right)^2 
+ \left(\Phi_2^\dagger\Phi_2\right) ^2,\\
{\rm Tr}\left[\left(H^\dagger H\right) \sigma_3\left(H^\dagger H\right)\sigma_3\right]  
&=& \left(\Phi_1^\dagger\Phi_1 - \Phi_2^\dagger\Phi_2\right)^2 + 2
\left(\Phi_1^\dagger\Phi_2\right) \left(\Phi_2^\dagger\Phi_1\right),\\
\det H^2 &=& \left(\Phi_2^\dagger\Phi_1\right)^2,\\
\det H^\dagger H &=& \left(\Phi_2^\dagger\Phi_1\right)\left(\Phi_1^\dagger\Phi_2\right).
\ee
The VEV of $H$ is expressed by a diagonal matrix 
\be
H = {\rm diag}(v_1,v_2).
\ee
For later purposes, let us discuss several simplified cases here. When we take 
\be
m_{11} = m_{22},\  \  \beta_1=\beta_2, 
\label{eq:simplify1}
\ee
the VEV of $H$ is proportional to the unit matrix, and the potential takes the following form:
\be
V_{v_1=v_2} 
&=&  -m^2 {\rm Tr}(H^\dagger H)
+ \lambda_1{\rm Tr}\left((H^\dagger H)^2\right) + \lambda_2 ({\rm Tr}(H^\dagger H))^2
+ \lambda_4 {\rm Tr}\left(H^\dagger H\sigma_3 H^\dagger H\sigma_3\right)
\nonumber\\
&-&  m_{12}^2 \left( \det H + {\rm h.c.}\right)
+  \left(\frac{\beta_5}{2}\det H^2 + {\rm h.c.}\right),
\ee
where we have used the Cayley--Hamilton identity 
${\rm Tr}[A^2] - ({\rm Tr}A)^2 + 2 {\rm det} A = 0$ for a two by two matrix $A$, and have defined new parameters by
\be
m_{11}^2 = -m^2,\quad
\lambda_1 = \frac{4\beta_1-3\beta_3-6\beta_4}{12},\quad
\lambda_2 = \frac{\beta_3 + \beta_4}{2},\quad
\lambda_4 = \frac{4\beta_1-3\beta_3}{12}.
\ee
If we further assume 
\be
\beta_1 = \frac{3}{4}\beta_3,
\label{eq:simplify2}
\ee
in addition to Eq.~(\ref{eq:simplify1}), terms with $\sigma_3$ disappears, in which case the potential manifestly has the custodial $SU(2)$ symmetry.
It should be noted here that, even when the potential has the custodial symmetry, it is explicitly broken by the existence of $U(1)_Y$ gauge interaction, which is the subgroup of $SU(2)_{R}$.
Therefore the custodial symmetry becomes the symmetry of whole Lagrangian only when we assume Eqs.~(\ref{eq:simplify1}), (\ref{eq:simplify2}) and turn off the $U(1)_Y$ gauge interaction.

It is also important to note that if we take
\be
m_{12} = \beta_5 = 0,
\label{eq:simplify3}
\ee
there is an additional $U(1)_a$ symmetry which rotates the relative phase of $\Phi_1$ and $\Phi_2$ as
\be
U(1)_a:\ H \to e^{i\alpha} H.
\label{eq:U1a}
\ee
Since this symmetry is spontaneously broken in the vacuum, the corresponding Nambu-Goldstone field appears.

When we discuss the existence of a non-Abelian string in the 2HDM 
in the following sections, 
we often consider the most symmetric Lagrangian with 
the following simplification,
\be
m_{11}=m_{22},\ \ \beta_1=\beta_2=\frac{3}{4}\beta_3,\ \ m_{12}=\beta_5=0,
\ee
namely all the simplifications, Eqs.~(\ref{eq:simplify1}), (\ref{eq:simplify2}) and (\ref{eq:simplify3}).
Under this assumption, the Higgs potential takes the following form with only three parameters:
\be
V = -m^2 {\rm Tr}[H^\dagger H] + \lambda_1 {\rm Tr}\!\left[(H^\dagger H)^2\right] 
+  \lambda_2 \left({\rm Tr}[H^\dagger H]\right)^2.
\label{eq:simplify_V}
\ee
For the moment, we also turn off the $U(1)_Y$ gauge coupling ($\sin\theta_W = 0$), then the global custodial SU(2) symmetry is exact. The vacuum of the $H$ field is expressed as
\be
\langle H \rangle = v{\bf 1}_2,\quad v^2 = \frac{m^2}{2\left(\lambda_1 + 2\lambda_2\right)},
\label{eq:VEV}
\ee
and the vacuum stability condition reads
\be
\lambda_1 + 2\lambda_2 > 0.
\label{eq:vac_c1}
\ee
In the vacuum, the symmetry is spontaneously broken as
\be
U(1)_a \times SU(2)_W \times SU(2)_{R} \to SU(2)_{C} 
\times (\mathbb{Z}_2)_{W+a}
.
\label{eq:sym_breaking}
\ee
Here, $(\mathbb{Z}_2)_{W+a}$ is defined as $(\mathbb{Z}_2)_{W+a}: (\omega {\bf 1}_2, {\bf 1}_2, \omega) \in SU(2)_W\times SU(2)_R \times U(1)_a$, where $\omega$ is defined by $\omega = e^{i\pi} = -1$.\footnote{
The discrete symmetry $(\mathbb{Z}_2)_{W+a}$ can be replaced by $(\mathbb{Z}_2)_{R+a}$ which
is defined by $(\mathbb{Z}_2)_{R+a}: ({\bf 1}_2, \omega {\bf 1}_2, \omega) \in SU(2)_W\times SU(2)_R \times U(1)_a$.
The difference between $(\mathbb{Z}_2)_{W+a}$ and $(\mathbb{Z}_2)_{R+a}$ can be absorbed by the center of $SU(2)_C$.
}
Therefore, the order parameter space is
\be
 M = \frac{U(1)_a \times SU(2)_W \times SU(2)_{R}}{SU(2)_{C}
\times (\mathbb{Z}_2)_{W+a}} \simeq \frac{U(1)\times SU(2)}{\mathbb{Z}_2} \simeq U(2).
\ee

Let us discuss masses of scalar bosons. Let us expand the fields $H$  
around the vacuum in terms of small fluctuations 
\be
H = v {\bf 1}_2 + \sigma_a h^a, \quad (\sigma_a = ({\bf 1}_2,\vec \sigma)),
\ee
where $h^a$ $(a=0,1,2,3)$ are complex scalar fields. 
Plugging this into Lagrangian, we find the mass spectrum: there are four NG bosons (three of them are
eaten by the weak gauge bosons) and  four massive modes. Those massive modes can be classified by the $SU(2)_{C}$ symmetry, and their masses are expressed as
\begin{eqnarray}
m_1^2 = 2m^2,\quad m_{\bf 3}^2 = \frac{\lambda_1}{\lambda_1 + 2\lambda_2}2m^2 = 4\lambda_1 v^2,
\end{eqnarray}
where $m_1$ is the mass of the $SU(2)_{C}$ singlet scalar (the real part of $h_0$) 
and $m_{\bf 3}$ is that of the adjoint components (the real parts of $h_{1,2,3}$).
From the above expressions, we see that the relation between $SU(2)_{C}$ singlet and triplet bosons becomes 
\be
m_1^2 \lesseqgtr m_{\bf 3}^2 \qquad \text{for}\quad \lambda_2 \lesseqgtr 0.
\ee  
The mass of the $SU(2)_W$ bosons is ($Z$ and $W$ are degenerate because of $\sin\theta_W = 0$)
\begin{eqnarray}
m_W^2 = g^2v^2.
\end{eqnarray}
Note that our $v^2$ is related to a conventional one by $v_{\rm EW}^2 = 4 v^2$.


\section{Topological non-Abelian strings at $\sin\theta_W = 0$}\label{sec:NAstring}

In this section, we discuss the existence of a non-Abelian string in the 2HDM.  
The symmetry breaking pattern in Eq.~(\ref{eq:sym_breaking}) is homotopically non-trivial:
\be
\pi_1\left(M\right) = \mathbb{Z}.
\ee
This will support the existence of several kinds of topologically stable strings as explained in the followings.

\subsection{Topological Abelian string}
Let us start from explaining an Abelian global string, which is characterized by the following form of field configurations:
\be
H = v f(r)e^{i\theta}{\bf 1}_2,\quad W^a_\mu = 0.
\ee
Here $(r,\theta)$ is the polar coordinate of the plane ($xy$ plane) perpendicular to the vortex string ($z$-axis).
The regularity requires $f(r) \to 0$ as $r\to0$, while $f(r) \to 1$ as $r \to \infty$ to minimize the energy of the system.
Since the $SU(2)_W$ gauge fields are vanishing everywhere, the vortex string  has no $SU(2)_W$ flux.
The topological charge of the string is given by $\pi_1(U(1)_a) = \mathbb{Z}$.

Dominant contribution to the tension of the global string comes 
from the kinetic term of $H$. One can see this by looking at the field configuration of $H$ at $r\to\infty$:
\be
H \to v e^{i\theta}{\bf 1}_2,
\qquad (r\to\infty).
\ee
Then, the energy density asymptotically ($r\to\infty$) behaves as
\be
{\cal E} \to {\rm Tr}\left[(\p_iH)^\dagger \p_i H\right]
\simeq 
2 \times \frac{v^2}{r^2} + \cdots,
\ee
where the factor $2$ comes from the trace. 
Integrating this over on the $x$-$y$ plane, we obtain the
tension (mass per unit length) of the string as
\be
T \simeq \int d^2x\ {\rm Tr}\left[(\p_i H)^\dagger \p_iH \right] \simeq 
4 \pi v^2 \log \Lambda + \cdots,
\ee
where $\Lambda$ is the IR cut-off parameter, and the ellipses stand for a finite contributions.

\subsection{Topological non-Abelian string}
As we will show shortly, the Abelian global string explained above is not the most elemental topological excitation
in the model. The most elemental string is called a non-Abelian string which is a partially global and partially local string. The ansatz of field configurations is taken as
\begin{eqnarray}
H_0 = v \left(
\begin{array}{cc}
h(r) & 0 \\
0 & f(r) e^{i\theta}
\end{array}
\right),\quad
W_{i,0}^a = \delta^{a3}\frac{1}{g}\epsilon_{ij}\frac{x^j}{r^2}(1-w(r)),
\quad W_{3,0}^a = 0,
\label{eq:rep}
\end{eqnarray}
with an appropriate boundary condition
\begin{eqnarray}
h'(0) = 0,\ f(0) = 0,\ w(0)=1,\quad h(\infty) = f(\infty) = 1,\ 
w(\infty) = 0.
\end{eqnarray}
Indication of the subscript $0$ in Eq.~(\ref{eq:rep}) will be tuned out to be clear below.
Note that $h(0)$ is not necessary zero because the left-top component does not wind. 
One may rewrite the above Higgs field configuration as
\be
H_0 = v e^{i\frac{\theta}{2}} e^{-i\frac{\theta}{2}\sigma_3}
\left(
\begin{array}{cc}
h(r) & 0\\
0 & f(r)
\end{array}
\right).
\ee
This shows that half of the phase appearing in the lower-right corner of $H$ in Eq.~(\ref{eq:rep}) is from the global $U(1)_a$ transformation, while the rest of the phase is supplied by $SU(2)_W$ rotation.
Therefore, this string has both features of a global string and a local string: 
Namely, its tension is logarithmically divergent while it carries the quantized magnetic flux along the vortex.

\begin{figure}[t]
\begin{center}
\includegraphics[width=15cm]{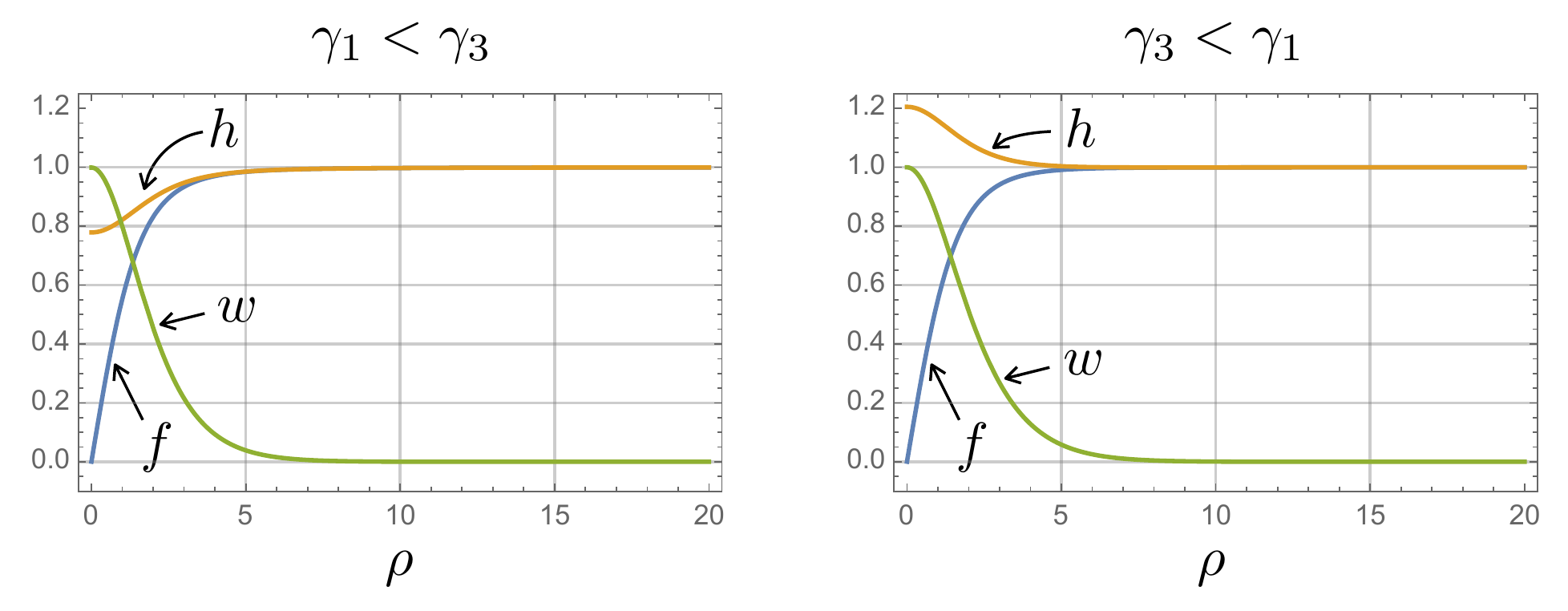}
\caption{The profile functions of $f(r)$, $h(r)$ and $w(r)$ for 
numerical solutions of an axially symmetric non-Abelian vortex string at the limit $g' \to 0$.
The left solution is for $(\gamma_1,\gamma_3)=(1,2)$  and the right one is for $(\gamma_1,\gamma_2)=(2,1)$.}
\label{fig:NA_vor_sinw_0}
\end{center}
\end{figure}

The energy density in terms of $f,h,w$ reads
\be
{\cal E} &=& \frac{v^2 m_W^2}{4} \bigg(
2 f'{}^2 + 2 h'{}^2 + \frac{w'{}^2}{\rho^2} + \frac{(w+1)^2}{2\rho^2}f^2 + \frac{(w-1)^2}{2\rho^2}h^2  \nonumber\\
&&- \gamma_1^2 (f^2+h^2)
+ \frac{\gamma_1^2+\gamma_3^2}{4}(f^4+h^4) + \frac{\gamma_1^2-\gamma_3^2}{2}f^2h^2
\bigg),
\ee
where $\rho = m_W r /\sqrt{2}$ 
and the prime stands for a derivative in terms of $\rho$. $\gamma_{1}$ and $\gamma_{3}$ are defined as dimensionless combinations of parameters as:
\be
\gamma_1 = \frac{\sqrt{2} m_1}{m_W},\quad \gamma_3 = \frac{\sqrt{2} m_3}{m_W}.
\ee
The equations of motion are summarized as
\begin{align}
&f'' + \frac{f'}{\rho} - \frac{(w+1)^2}{4\rho^2}f-\frac{1}{4}\left[\left(\gamma_1^2 + \gamma_3^2\right)f^2 + 
\left(\gamma_1^2-\gamma_3^2\right)h^2-2\gamma_1^2\right] f = 0,
\label{eq:f1}\\
&h'' + \frac{h'}{\rho} - \frac{(w-1)^2}{4\rho^2}h-\frac{1}{4}\left[\left(\gamma_1^2 - \gamma_3^2\right)f^2 + 
\left(\gamma_1^2+\gamma_3^2\right)h^2-2\gamma_1^2\right] h = 0,
\label{eq:h1}\\
&w''-\frac{w'}{\rho} - \frac{w+1}{2}f^2-\frac{w-1}{2}h^2 = 0.
\label{eq:w1}
\end{align}
Note that only two dimensionless parameters,  $\gamma_{1}$ and $\gamma_{3}$, appear in the above expressions.
In Fig.~\ref{fig:NA_vor_sinw_0}, we show profile functions that are obtained by solving above equations for the cases of $(\gamma_1, \gamma_3) = (1, 2)$ (left panel) and $(2, 1)$ (right panel). Shapes of $f$ and $w$ are quite similar for the both cases, while $h$ behaves in a qualitatively different way. It is a general tendency that $h$ becomes smaller (larger) than one near the vortex core in the case of $\gamma_1 < \gamma_3$ ($\gamma_3 < \gamma_1$).

In contrast to the purely global solution, the $SU(2)_W$ gauge fields take non-trivial configurations.
The $SU(2)_W$ magnetic flux is given by
\begin{eqnarray}
W^3_{12} = \frac{w'(r)}{gr},\quad
\Phi^3_{12} = \int d^2x\ W^3_{12} = -\frac{2\pi}{g}.
\end{eqnarray}

\begin{figure}[t]
\begin{center}
\includegraphics[width=8cm]{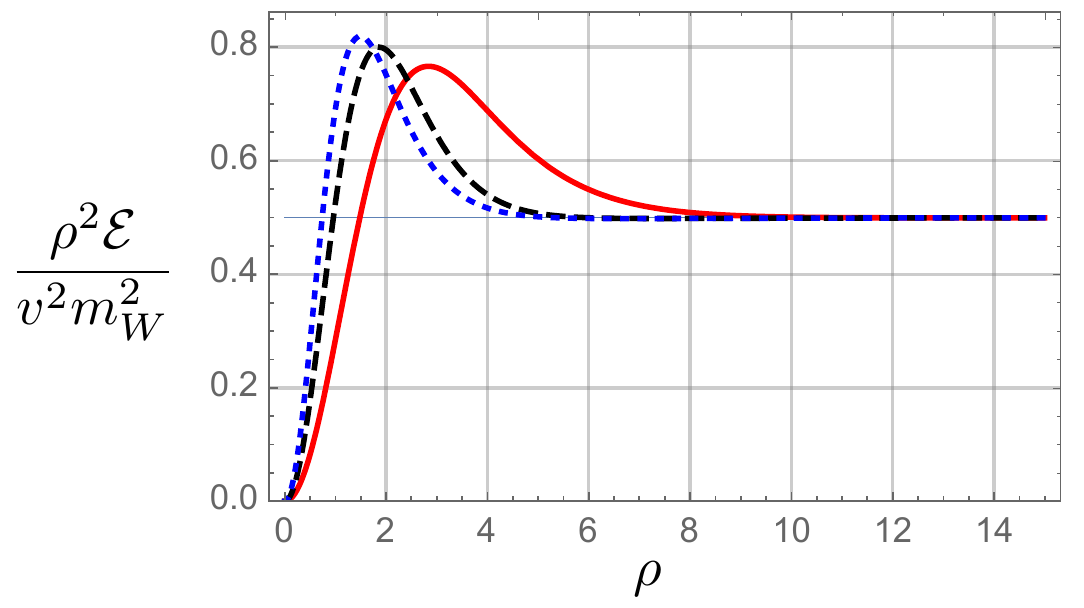}
\caption{$\rho^2 {\cal E}/v^2$ for the non-Abelian strings with 
$(\gamma_1,\gamma_3) = (1,\frac{1}{2})$ (red solid curve) and $(1,1)$ (black dashed curve),
and $(1,2)$ (blue dotted curve).
All the curves converge
to $1/2$, indicating the fractional $U(1)_a$ winding number, as expected from Eq.~(\protect\ref{eq:energy_density}).
}
\label{fig:fig2a}
\end{center}
\end{figure}

Similarly to the Abelian global string, dominant contribution to the tension of a non-Abelian string 
comes from the kinetic term of $H$ at $r\to\infty$.
The asymptotic behavior of a non-Abelian string reads
\be
H \to v e^{i\frac{\theta}{2}}{\bf 1}_2,
\qquad (r\to\infty),
\label{eq:Hasy}
\ee
where we have gauged away irrelevant phase.
Then, the energy density asymptotically ($r\to\infty$) behaves as
\be
{\cal E} \sim {\rm Tr}\left[(\p_iH)^\dagger \p_i H\right] \to
2 \times \frac{1}{2^2} \times \frac{v^2}{r^2} + \cdots,
\label{eq:energy_density}
\ee
where the factor $\frac{1}{2^2}$ reflects the fact 
that the $U(1)_a$ winding number is $1/2$. 
This asymptotic behavior is well reproduced by numerical solutions irrespective of the choice of parameters
$\gamma_1,\gamma_2$ as shown in Fig.~\ref{fig:fig2a}.
Integrating the energy density (\ref{eq:energy_density}) over $x$-$y$ plane, we obtain the
tension (mass per unit length) of the string as
\be
T \simeq \int d^2x\ {\rm Tr}\left[(\p_i H)^\dagger \p_iH \right] \simeq 
\pi v^2 \log \Lambda + \cdots,
\label{eq:asym_H}
\ee
where $\Lambda$ is the IR cut-off parameter, and the ellipses stand for finite contributions.
It is now clear that the tension of the non-Abelian string is about one quarter of that of 
the Abelian global string. Therefore, the former is considered to be more fundamental topological excitation compared to the latter. One should note that a single Abelian global vortex and a system which has two non-Abelian vortices are in the same topological sector since 
the $U(1)_a$ winding number of the Abelian global string is twice of that of the non-Abelian string. Therefore, those two systems are related by continuous deformation of field configurations, and an Abelian global string, even if it is  created, is likely to decay into two non-Abelian strings since well-separated two non-Abelian strings are energetically favored compared to a single Abelian global string.

\begin{figure}[t]
\begin{center}
\includegraphics[width=15cm]{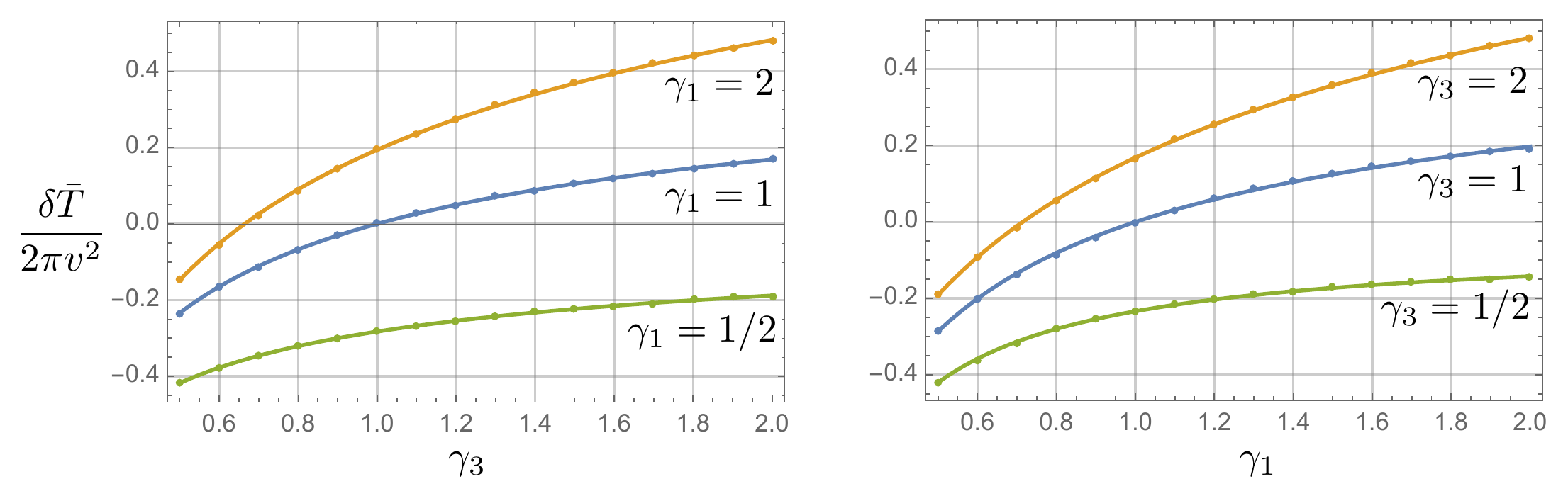}
\caption{The $\gamma_3$ ($\gamma_1$) dependence of the finite contribution $\delta \bar T/2\pi v^2$ is shown for
the three different $\gamma_1 = \frac{1}{2},1,2$ ($\gamma_3 = \frac{1}{2},1,2$) in the left (right) panel.  
}
\label{fig:fig2b}
\end{center}
\end{figure}

Now let us discuss a parameter dependence of the tension of the non-Abelian strings. For the purpose of removing the logarithmic divergence from the discussion, we take the tension of the string for the case of $(\gamma_1, \gamma_3) = (1, 1)$ as a reference, and defined the following quantity for general values of $(\gamma_1, \gamma_3) $:
\be
\delta \bar T(\gamma_1,\gamma_3) = 2\pi v^2 \int_0^{\infty}  d\rho\, \rho \left[{\cal E}(\rho;\gamma_1,\gamma_3) - 
{\cal E}(\rho;\gamma_1=1,\gamma_3=1)\right].
\ee
Fig.~\ref{fig:fig2b} shows the $\gamma_3$ ($\gamma_1$) dependence of $\delta \bar T$  for three choices of the value of $\gamma_1$ ($\gamma_3)$.
From the figure, one can see that $\delta \bar T$ monotonically increases as a function of $\gamma_3$ and $\gamma_1$.

One of distinctive features of the non-Abelian strings is the existence of non-Abelian zero modes. Since the presence of a non-Abelian string spontaneously breaks the $SU(2)_{C}$ custodial symmetry, those zero modes appear as NG modes.  As we showed in Eq.~(\ref{eq:Hasy}), while the vortex configuration asymptotically approaches to an $SU(2)_{C}$-conserving form, the $SU(2)_{C}$ symmetry is spontaneously broken at the center of the string by the following form of configuration:
\be
H\big|_{\text{NA string}} \to v e^{i\frac{\theta}{2}} \left(
\begin{array}{cc}
h(0) & 0 \\
0 & 0
\end{array}
\right),
\qquad (r\to0).
\ee
Therefore the non-Abelian NG modes are localized near the string where  the $SU(2)_{C}$ symmetry is broken spontaneously. 
The presence of the string configuration breaks the $SU(2)_{C}$ symmetry  down to $U(1)_{c}$, generating the non-Abelian NG modes on the coset of
\be
\frac{SU(2)_{C}}{U(1)_c} \simeq \mathbb{C}P^1 \simeq S^2.
\ee
Such a coset manifold is called the moduli space and its coordinates are called the moduli parameters. Different points on the $S^2$ moduli space correspond to physically different degenerate string solutions.  We identify the solution given in Eq.~(\ref{eq:rep}) as the one associated with 
the north pole of  $S^2$ moduli space, and we call it a $(0,1)$-string. 

The $(1,0)$-string corresponds to the antipodal point, namely the south pole, which is given by
\begin{eqnarray}
H_{\pi} = v \left(
\begin{array}{cc}
f(r)e^{i\theta} & 0 \\
0 & h(r)
\end{array}
\right),\quad
W_{i,\pi}^a = -\delta^{a3}\frac{1}{g}\epsilon_{ij}\frac{x^j}{r^2}(1-w(r)),
\quad W_3^a = 0,
\label{eq:rep2}
\end{eqnarray}
$W^a_i$ of the $(1, 0)$-string solution has the same form as that of the $(0,1)$-string solution with an opposite sign. Therefore the $SU(2)_W$ magnetic flux of the $(1, 0)$-string takes the same magnitude as that of a $(0,1)$-string with the opposite sign: 
\begin{eqnarray}
W^3_{12} = -\frac{w'(r)}{gr},\quad
\Phi^3_{12} = \int d^2x\ W^3_{12} = \frac{2\pi}{g}.
\end{eqnarray}

String solutions on a generic point of the $S^2$ moduli space can be obtained by acting an $SU(2)_{C}$ transformation on the $(0,1)$-string configuration, given in Eq.~(\ref{eq:rep}).
We first apply a rotation around the $\sigma_2$-axis which is  
represented by the following $SU(2)_{C}$ transformation:
\be
U(\zeta) = \exp\left(\frac{i\sigma_2}{2}\zeta\right) = 
\left(
\begin{array}{cc}
\cos\frac{\zeta}{2} & \sin\frac{\zeta}{2}\\
-\sin\frac{\zeta}{2} & \cos\frac{\zeta}{2}
\end{array}
\right),\quad 0 \le \zeta \le \pi.
\label{eq:U}
\ee
\begin{figure}[t]
\begin{center}
\includegraphics[width=6cm]{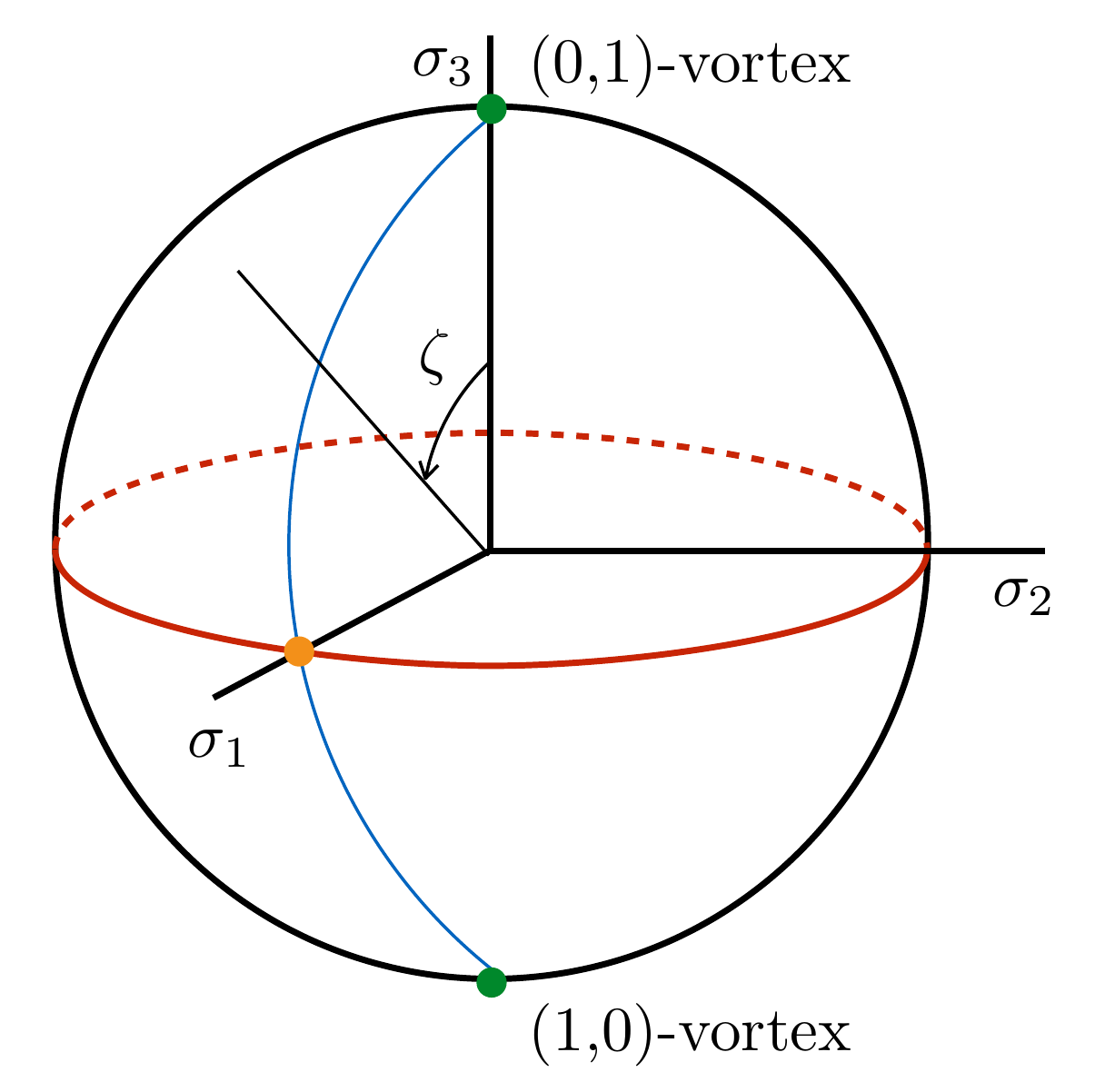}
\caption{The moduli space $S^2$ of a non-Abelian vortex string. }
\label{fig:moduli}
\end{center}
\end{figure}
Acting this on $H$ and $W_i$ of a $(0,1)$-string, both of which are the adjoint representation of $SU(2)_{C}$, we have
\be
H_\zeta &=& U H U^\dagger = v e^{i\frac{\theta}{2}}\left[
\frac{g e^{-i\frac{\theta}{2}}+f e^{i\frac{\theta}{2}}}{2} {\bf 1}_2 + \frac{g e^{-i\frac{\theta}{2}} - f e^{i\frac{\theta}{2}}}{2}
\sigma_\zeta\right],\label{eq:Htheta}\\
W_{i,\zeta} &=& U W_i U^\dagger =
\frac{1}{g}\epsilon_{ij}\frac{x^j}{r^2}\left(1 - w(r)\right)\, \sigma_\zeta,
\label{eq:Wtheta}
\ee
with
\be
\sigma_\zeta \equiv U \sigma_3 U^\dagger = \sigma_3\cos\zeta - \sigma_1\sin\zeta.
\ee
Generic solution can be obtained by further applying rotation around  the $\sigma_3$-axis in $SU(2)_{C}$. 
The schematic
picture of the moduli space is given in Fig.~\ref{fig:moduli}.

\subsection{Non-topological local non-Abelian string}

There is the third type of string solution which takes the configuration on purely $SU(2)_W$ gauge orbits.
The simplest ansatz is given by
\be
H = f(r)e^{-\frac{i\sigma_3}{2}2\theta} \left(\begin{array}{cc}
v & 0\\
0 & v
\end{array}
\right),\quad
W_i^3 = \frac{2}{g}\epsilon_{ij}\frac{x^j}{r^2}\left(1 - w(r)\right).
\ee
This is clearly local string, and therefore the tension is finite. 
However, since the $SU(2)_W$ has
topologically trivial first homotopy group $\pi_1(SU(2)_W) = \{\phi\}$, 
the stability of the string is not topologically ensured.

\section{Topological $Z$- and $W$-strings for $\sin\theta_W \neq 0$  at $\tan\beta=1$}
\label{sec:U(1)Y}

Let us next turn on the $U(1)_Y$ gauge coupling and see how the non-Abelian string solution is affected. With the existence of the $U(1)_Y$ gauge interaction, the custodial symmetry is no longer a symmetry of the model. 
As a consequence, NG modes localized on a string obtain mass to become pseudo-NG modes, and 
almost all the points of the $S^2$ moduli space are energetically lifted, 
that is, almost all the non-Abelian string solutions become unstable 
leaving only two exceptions; the $(1,0)$- and $(0,1)$-string remain stable solutions even after switching on the $U(1)_Y$ gauge interaction. 

For illustration, let us again consider the simplified model with
the Higgs potential given in Eq.~(\ref{eq:simplify_V}).
The VEV is the same as before, namely $H = v {\bf 1}_2$. Therefore, the model has $\tan\beta=v_2/v_1 =1$.
The $U(1)_Y$ and $\sigma_3$ parts of the $SU(2)_W$ are mixed in the same way as those in the SM. The mass eigenstates, namely a massive $Z$ boson and massless photon, are expressed as follows:
\be
Z_\mu = \cos\theta_W\, W^3_\mu  \, - \, \sin\theta_W\, B_\mu,\\
A_\mu = \sin\theta_W\ W_\mu^3 \, + \, \cos\theta_W\, B_\mu.
\ee
Here, the mixing angle $\theta_W$ is defined by $\cos\theta_W = g/\sqrt{g^2+g'^2}$.

\subsection{Topological $Z$-string}
The $(1,0)$- and $(0,1)$-strings are the topologically stable $Z$-string \cite{Dvali:1993sg}.
To describe it in details, let us first set $W_\mu^{\pm} = A_\mu = 0$ since these fields do not play any role for string solutions.
Then the covariant derivative of the diagonal Higgs field reads
\be
D_\mu H = \left(\p_\mu - \frac{g}{\cos\theta_W}\,\frac{i\sigma_3}{2}Z_\mu\right)H.
\label{eq:DZ}
\ee
Note that difference from the case of $\sin\theta_W = 0$ in the last section 
is just $W_\mu^3 \to Z_\mu$ and $g \to g/\cos\theta_W$.
Therefore, the ansataz for the configuration of a $(0,1)$-string is obtained from Eq.~(\ref{eq:rep}) by simply replacing the coupling as follows:
\begin{eqnarray}
H_0 = v \left(
\begin{array}{cc}
h(r) & 0 \\
0 & f(r) e^{i\theta}
\end{array}
\right),\quad
Z_{i,0} = \frac{g}{\cos\theta_W}\epsilon_{ij}\frac{x^j}{r^2}(1-w(r)).
\label{eq:Z_string}
\end{eqnarray}
In terms of the $SU(2)_W$ and $U(1)_Y$ gauge fields, these are expressed as
\be
W^3_{i,0} =  \frac{g}{g^2+g'{}^2}\epsilon_{ij}\frac{x^j}{r^2}(1-w(r)) ,\quad
B_{i,0} =  -\frac{g'}{g^2+g'{}^2}\epsilon_{ij}\frac{x^j}{r^2}(1-w(r)).
\label{eq:Z_string2}
\ee
The equations of motion for the profile functions $f,h,w$ are given by
\begin{align}
&f'' + \frac{f'}{\tilde\rho} - \frac{(w+1)^2}{4\tilde\rho^2}f-\frac{1}{4}\left[\left(\tilde\gamma_1^2 + \tilde\gamma_3^2\right)f^2 + 
\left(\tilde\gamma_1^2-\tilde\gamma_3^2\right)h^2-2\tilde\gamma_1^2\right] f = 0,\\
&h'' + \frac{h'}{\tilde\rho} - \frac{(w-1)^2}{4\tilde\rho^2}h-\frac{1}{4}\left[\left(\tilde\gamma_1^2 - \tilde\gamma_3^2\right)f^2 + 
\left(\tilde\gamma_1^2+\tilde\gamma_3^2\right)h^2-2\tilde\gamma_1^2\right] h = 0,\\
&w''-\frac{w'}{\tilde\rho} - \frac{w+1}{2}f^2-\frac{w-1}{2}h^2 = 0.
\end{align}
where $\tilde \rho = m_Z r /\sqrt2$, $\tilde\gamma_{1,3} = \sqrt2 m_{1,3}/m_Z$ with
\be
m_Z^2 = \frac{g^2v^2}{\cos^2\theta_W}.
\ee
The dominant part of the string tension is independent of $\sin\theta_W$ since
the logarithmic divergence comes from the asymptotic behavior $H \sim e^{\frac{i\theta}{2}} v{\bf 1}_2$. 
Indeed, the dominant part reads
\be
T_Z = \pi v^2 \log \Lambda + \cdots,
\label{eq:TZ}
\ee 
as was given
in Eq.~(\ref{eq:asym_H}).
A difference between the $Z$-string for $\sin\theta_W \neq 0$ and the non-Abelian string with $\sin\theta_W = 0$ appears in the subdominant finite part. 
The configuration of as $(1,0)$-string is obtained from that of a $(0,1)$-string by changing the sign of gauge fields. Tensions of $(1,0)$- and $(0,1)$- strings are exactly the same.
Near the center of a $Z$-string, the $Z$-flux with the following magnitudes is confined:
\be
\Phi_Z  = \pm \frac{2\pi \cos\theta_W}{g}.
\ee
Here the plus (minus) sign is for the $(1,0)$-string ($(0,1)$-string).

\subsection{Explicitly broken $SU(2)_{C}$ transformation}

Now, let us discuss what happens when we transform the $Z$-string solution by $SU(2)_{C}$ rotations. We first note that although the $SU(2)_{C}$ rotation is not the symmetry of the Lagrangian, the vacuum is invariant under the $SU(2)_{C}$ rotation since we are now considering the situation that the vacuum expectation value of $H$ is proportional to identity: $\langle H \rangle = v {\bf 1}_2$ ($\tan\beta=1$).
Therefore, the asymptotic form of the $Z$-string configuration are not affected by the $SU(2)_{C}$ rotations since they approaches to the vacuum state at $r \to \infty$. This means that the logarithmic divergence of the string tension, which originates from large $r$ behavior of the string configuration, is not changed by an $SU(2)_{C}$ rotation. Meanwhile, the finite part of the string tension is affected by  $SU(2)_{C}$ rotations since string configurations near the center of the vortex are not invariant under those rotations. We will show how the string tension are affected by $SU(2)_{C}$ term later in this section

\subsection{Topological $W$-string}

Before investigating the general case, let us examine a special string solution that are obtained by applying an $SU(2)_{C}$ rotation with $\zeta = \pi /2$ (see Eq.~(\ref{eq:U})) to the $Z$-string solution.
This corresponds to a point on the equator of the $S^2$ space shown in Fig.~\ref{fig:moduli}.
The string constructed in this way has an axis which is orthogonal to that of the $Z$-string, and the vortex is made of $H$ and $W^\pm$. Therefore we call this string as a topological $W$-string. Since the existence of $U(1)_Y$ is irrelevant for the $W$-string, 
the corresponding solution is precisely the same as that in the case of $\sin\theta_W = 0$.
Hence, the ansatz is the same as those 
in Eqs.~(\ref{eq:Htheta}) and (\ref{eq:Wtheta}) with $\zeta = \frac{\pi}{2}$:
\be
H &=& \frac{v}{2}e^{i\frac{\theta}{2}} \left(
\begin{array}{cc}
f e^{i\frac{\theta}{2}} + h e^{-i\frac{\theta}{2}} & 
f e^{i\frac{\theta}{2}} - h e^{-i\frac{\theta}{2}}\\
f e^{i\frac{\theta}{2}} - h e^{-i\frac{\theta}{2}} & 
f e^{i\frac{\theta}{2}} + h e^{-i\frac{\theta}{2}}
\end{array}
\right) \xrightarrow{r\to\infty} v e^{i\frac{\theta}{2}}
\left(
\begin{array}{cc}
\cos \frac{\theta}{2} & i \sin\frac{\theta}{2}\\
i \sin\frac{\theta}{2} & \cos\frac{\theta}{2}
\end{array}
\right) = v e^{i\frac{\theta}{2}} e^{i\sigma_1\frac{\theta}{2}},\nonumber\\
W^1_i &=& -\frac{1}{g}\epsilon_{ij}\frac{x^j}{r^2}\left(1 - w(r)\right) \sigma_1
\xrightarrow{r\to\infty} -\frac{1}{g}\epsilon_{ij}\frac{x^j}{r^2} \sigma_1 = \frac{\p_i\theta}{g}\sigma_1.
\ee
The covariant derivative of $H$ asymptotically behaves as
\be
D_i H 
\to v \left(\p_i - \frac{ig}{2}\frac{\p_i\theta}{g}\sigma_1\right)e^{i\frac{\theta}{2}} e^{i\sigma_1\frac{\theta}{2}}
= \frac{v}{2}\, i\p_i\theta\, e^{i\frac{\theta}{2}}\, e^{i\sigma_1\frac{\theta}{2}}.
\ee
Therefore, the tension can be derived as  
\be
T_W = \pi v^2 \log \Lambda + \cdots,
\label{eq:TW}
\ee
which has the same form of the divergence with the same coefficient as 
the case of the $Z$-string, given in Eq.~(\ref{eq:TZ}). 
The $SU(2)_W$ flux confined in the string is given as 
\be
\Phi_W^1 = \frac{2\pi}{g}.
\ee
Thus, the $W$ flux quantum is different from the $Z$ flux quantum by the factor $\cos\theta_W$.

Note that there is one parameter (azimuthal angle) family of the topological $W$-strings corresponding to points on the equator of the $S^2$ space.
Since shift on the equator is generated by gauged $U(1) \in SU(2)_{C}$, all the points 
on the equator are physically equivalent. However, if there are more than one $W$-strings with different azimuthal angles, the relative difference remains physical.

\subsection{General unstable non-Abelian strings}

We have studied the topological $Z$- and $W$-strings in the case of $\sin\theta_W \neq 0$. A natural question to be asked is: Which is energetically favored? 
As we mentioned above, the tension of topological $W$-string is same as that of the non-Abelian string (with $\sin\theta_W =0$). Meanwhile, the tension of $Z$-string is obtained from the tension of the non-Abelian string by replacing $g$ with $\sqrt{g^2 + g'{}^2}$. The effect of the change of the coupling can be understood from Fig.~\ref{fig:fig2b} considering the fact that, for the tension of the non-Abelian string, changing the coupling with fixing all the other parameters is equivalent to changing $\gamma_{1,3}$  with fixing the value of the coupling. 
In Fig.~\ref{fig:fig2b}, we have seen that the string tension
monotonically increases as functions of $\gamma_1$ and $\gamma_3$. Since $\gamma_{1,3} \propto g^{-1}$, this
implies that the string for the larger $g$ has the smaller tension. Therefore, we conclude that the $Z$-string is always energetically
favored to the $W$-string in the simplest Higgs potential given in Eq.~(\ref{eq:simplify_V}).

To see this more clearly, 
let us now derive an ``effective potential" $V_{\rm eff}(\zeta)$, namely the string tension of configurations made of $SU(2)_{C}$ rotations from the $Z$-string, in an arbitrary point of the moduli space. 
As long as $g' \ll g$ holds, effects of gauging $U(1)_Y$ should be small. Therefore,
we can perturbatively deal with the effect around the solution with $g'=0$.
In order to connect smoothly the $Z$-strings (the north and south poles of $S^2$)
and a $W$-string (a point on the equator), 
we take a variational ansataz by rotating $H_0$ and $W_{i,0}$ given in Eqs.~(\ref{eq:Z_string}) and (\ref{eq:Z_string2}) (the ansatz for $\zeta = 0$)
by $U_\zeta \in SU(2)_{C}$ given in Eq.~(\ref{eq:U}).  On the other hand, we leave $B_{i,0}$ as it is in Eq.~(\ref{eq:Z_string2}).
We plug this $SU(2)_{C}$ rotated ansatz into the Lagrangian with the simplified potential (\ref{eq:simplify_V}),
and minimize energy for a fixed $\zeta$. Then, we repeat the procedure by varying $\zeta$ from 0 to $\pi$.
This continuously connects the $Z$- and $W$-strings. 

For practical use, we invert the above procedure as follows.
Since the gauge kinetic terms and the potential term are $SU(2)_{C}$ symmetric, 
the Higgs kinetic term is unique term which breaks the $SU(2)_{C}$ symmetry 
via the minimal coupling with the $U(1)_Y$ gauge field:
\be
D_i H_\zeta = U_\zeta\left(\p_i H_0 - g\frac{i}{2}W_{i,0}^3\sigma_3 H_0 
+ g'\frac{i}{2}H_0 \sigma_{-\zeta} B_i\right) U_\zeta^\dagger,
\label{eq:rotcov}
\ee
where $H_0$ takes the form given in Eq.~(\ref{eq:Z_string}).
Then, just for simplicity, we make a slightly different ansatz for $W_i$ and $B_i$ from Eq.~(\ref{eq:Z_string2}) as
\be
W_{i,0} = -\frac{1}{g}\epsilon_{ij}\frac{x^j}{r^2}(1-w(r)),\quad
B_{i,0} = -\frac{1}{g'}\epsilon_{ij}\frac{x^j}{r^2}b(r).
\label{eq:ansatz_B_W}
\ee
\begin{figure}[t]
\begin{center}
\includegraphics[width=10cm]{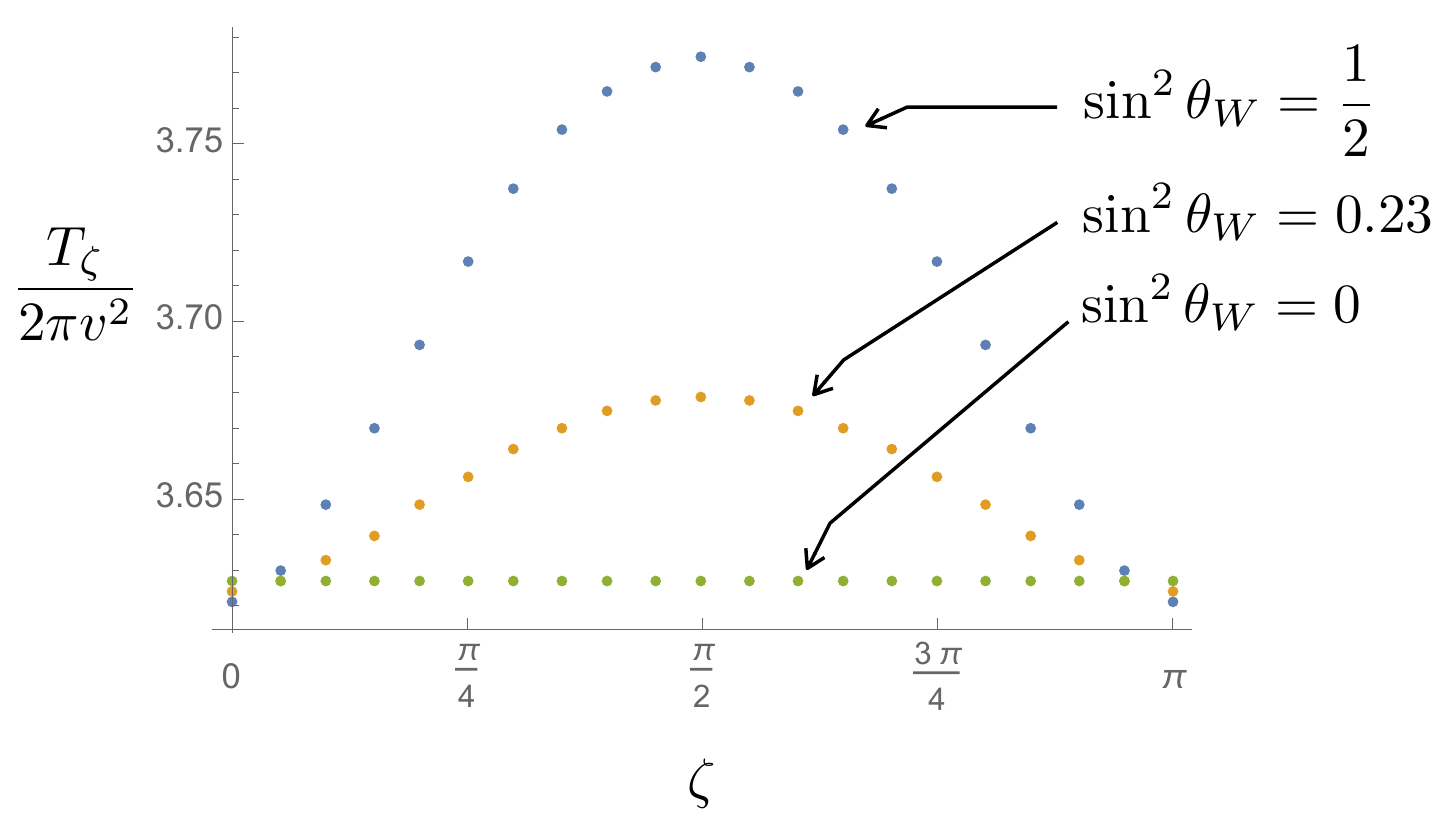}
\caption{The $\zeta$ dependence of the string tensions. We plot three cases of $\sin^2\theta_W = \frac{1}{2}, 0.23$, and $0$:
namely, $\frac{m_W^2}{m_Z^2} = \cos^2\theta_W = \frac{1}{2}, \left(\frac{80}{91}\right)^2$, and $1$. We fix the other parameter by requiring 
$\frac{\tilde\gamma_1}{\sqrt2} = \frac{m_1}{m_Z} = \frac{125}{91}$ and $m_3 = 100 m_1$.}
\label{fig:eff_pot_Y}
\end{center}
\end{figure}
The boundary condition for $w(r)$ and $b(r)$ should be chosen in such a way that the logarithmic divergences
given in Eqs.~(\ref{eq:TZ}) and (\ref{eq:TW}) holds regardless of $\zeta$.
The equations that minimizes the energy for a fixed value of $\zeta$ read
\begin{align}
&f'' + \frac{f'}{\tilde\rho} 
-\frac{\left((1+w)^2 - 2 (1+w)b \cos \zeta +b^2\right)}{4 \tilde\rho^2}f\nonumber\\
&\ -\frac{1}{4}\left[\left(\tilde\gamma_1^2 + \tilde\gamma_3^2\right)f^2 + 
\left(\tilde\gamma_1^2-\tilde\gamma_3^2\right)h^2-2\tilde\gamma_1^2\right] f = 0,
\label{eq:f1}\\
&h'' + \frac{h'}{\tilde\rho} 
-\frac{\left((1-w)^2 + 2 (1-w)b \cos\zeta +b^2\right)}{4 \tilde\rho^2}h\nonumber\\
&\ -\frac{1}{4}\left[\left(\tilde\gamma_1^2 - \tilde\gamma_3^2\right)f^2 + 
\left(\tilde\gamma_1^2+\tilde\gamma_3^2\right)h^2-2\tilde\gamma_1^2\right] h = 0,
\label{eq:h1}\\
&w''-\frac{w'}{\tilde\rho} 
- \frac{1}{2} \cos^2\theta_W \left( (1+w - b\cos\zeta)f^2 -  (1-w + b\cos \zeta)h^2 \right) = 0,
\label{eq:w1}\\
&b''-\frac{b'}{\tilde\rho} 
+ \frac{1}{2} \sin^2\theta_W \left( ((1+w)\cos\zeta - b)f^2 - ((1-w)\cos \zeta + b)h^2 \right) = 0.
\label{eq:b1}
\end{align}
We numerically solve these by varying $\zeta$ from $0$ to $\pi$ for a fixed $\theta_W$, and evaluate
the tension by integrating the energy density between $0 \le \tilde\rho \le 40$.
Fig.~\ref{fig:eff_pot_Y} shows the tension for each $\zeta$. Note that the value of the vertical axis in the figure is rather irrelevant since it logarithmically depends on the numerical cutoff of $\tilde{\rho}$ integration. What is important here is the difference of tensions among solutions for different input values of $\zeta$ and $\sin \theta_W$.
The edges ($\zeta = 0$ and $\pi$), corresponding to the $Z$-strings, are minima of the effective potential while the $W$-string is the maximum. 
This clearly shows that two $Z$-strings are most stable with degenerate and minimum tension and 
$W$-strings  having the largest tension,
corresponding to the maximum of the potential.


\section{
Topological strings with an $SU(2)_{C}$ breaking potential at $\tan\beta=1$}
\label{sec:mod-Higgs}

\subsection{The mass ordering of the Higgs}

\begin{figure}[t]
\begin{center}
\includegraphics[height=7cm]{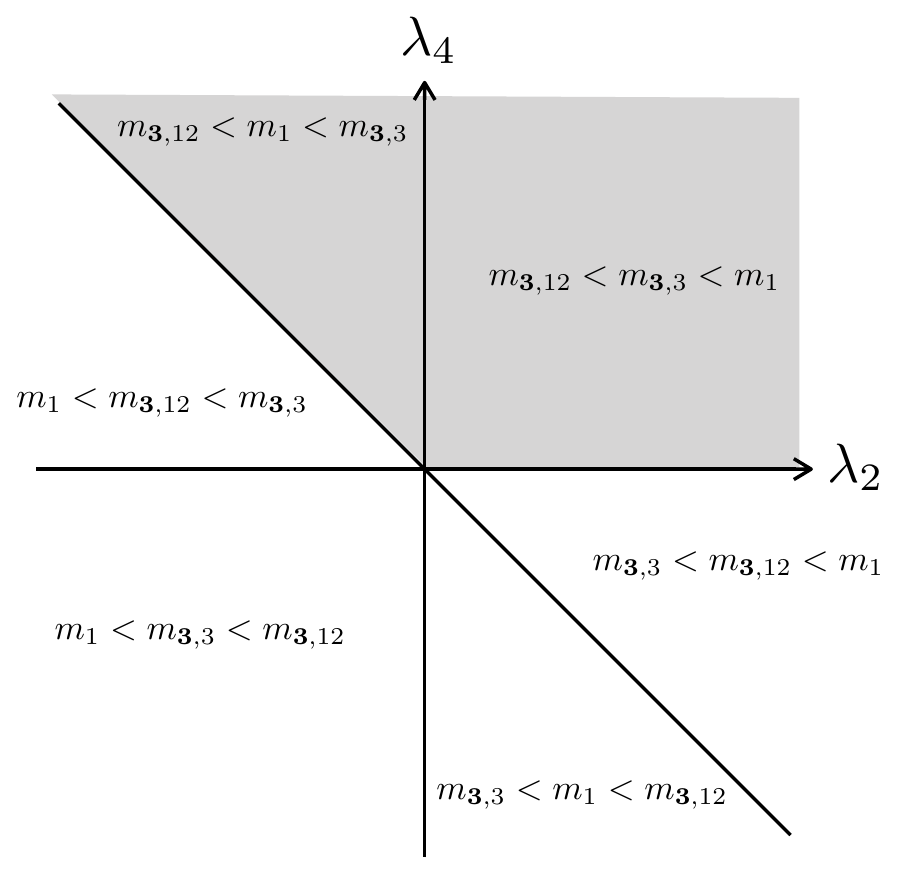}
\caption{The six possible mass orderings in terms of $\lambda_2$ and $\lambda_4$. The gray region is phenomenologically disfavored
since the lightest scalar field is not a neutral singlet scalar.}
\label{fig:lambda_2_4}
\end{center}
\end{figure}

We now add a term which explicitly breaks $SU(2)_{C}$ symmetry 
to the simplest potential given in Eq.~(\ref{eq:simplify_V}):
\be
V = -m^2 {\rm Tr}[H^\dagger H] + \lambda_1 {\rm Tr}\!\left[(H^\dagger H)^2\right] 
+  \lambda_2 \left({\rm Tr}[H^\dagger H]\right)^2
+ \lambda_4 {\rm Tr}[H^\dagger H \sigma_3 H^\dagger H \sigma_3].\nonumber\\
\label{eq:modified_V1}
\ee
The VEV of $H$ is still proportional to the identity matrix, however
$v$ given in Eq.~(\ref{eq:VEV}) is modified as
\be
H = v{\bf 1}_2,\quad v^2 = \frac{m^2}{2\left(\lambda_1 + \lambda_4 + 2\lambda_2\right)}.
\ee
The mass of the $SU(2)_{C}$ singlet scalar (the real part of $h_0$) 
is unchanged as
\be
m_1^2 = 2m^2,
\ee
whereas that of the $SU(2)_{C}$ triplet split into two folds as
\be
m_{{\bf 3},3}^2 = \frac{\lambda_1 + \lambda_4}{\lambda_1 + \lambda_4 + 2 \lambda_2} 2m^2,\quad
m_{{\bf 3},12}^2 = \frac{\lambda_1 - \lambda_4}{\lambda_1 + \lambda_4 + 2 \lambda_2} 2m^2,
\ee
where $m_{{\bf 3},3}$ is the mass of the real part of $h_3$, and 
$m_{{\bf 3},12}$ is the mass of the real parts of $h_1$ and $h_2$ 
(which would-be charged under $U(1)_{\rm EM}$) as a result of turning on $\lambda_4$.
Then, mass squares of the neutral scalars read
$m_1^2$ and $m_{{\bf 3},3}^2$. 
Depending on the parameters, there are six patterns of mass ordering.
We summarize those on the $\lambda_2$-$\lambda_4$ plane 
in Fig.~\ref{fig:lambda_2_4} ($\lambda_1$ should be set to keep $\lambda_1 + \lambda_4 + 2\lambda_2 > 0$ for the vacuum stability).

For phenomenological viability, we should identify the physical Higgs $h$ as the lightest neutral singlet scalar field. Therefore, we identify masses of SM-like Higgs ($m_h$) and that of heavy Higgs ($m_H$) as follows:
\be
\lambda_2 < 0&:&\quad m_h \equiv m_1,\quad m_H \equiv m_{{\bf 3},3},\\
\lambda_2 > 0&:&\quad m_h \equiv m_{{\bf 3},3},\quad m_H \equiv m_1.
\ee
On the other hand, $m_{{\bf 3},12}$ is always identified as the mass of the charged scalar $H^\pm$:
\be
m_{H^\pm} \equiv m_{{\bf 3},12}.
\ee
We exclude the parameter region where $m_{H^\pm}$ is the lightest scalar, inconsistent with phenomenology, which corresponds to the gray region of Fig.~\ref{fig:lambda_2_4}.

\subsection{Non-Abelian string at $\sin\theta_W = 0$}

The newly added $\lambda_4$ term breaks the $SU(2)_{C}$ symmetry explicitly to its $U(1)$ subgroup. 
Therefore, the degeneracy of the orientational moduli
$S^2$ space of non-Abelian string is resolved in a similar way when we switched on the $U(1)_Y$ gauge interaction.
Here, to see an effect of the $\lambda_4$ term, we first turn off the $U(1)_Y$ gauge coupling $g'$ ($\sin\theta_W= 0$).
Then, we derive the $\zeta$ dependence of the string tension in a similar manner that we did in Sec.~\ref{sec:U(1)Y}.
Namely, we take the same ansatzs in Eqs.~(\ref{eq:Htheta}), (\ref{eq:Wtheta}), and plug those into the Lagrangian. Then we obtain the following restricted equations of motion
for the profile functions
\begin{align}
&f'' + \frac{f'}{\rho} - \frac{(w+1)^2}{\rho^2}f 
-\frac{1}{8}\bigg[f^2 \left(2 \gamma_1^2 + \gamma_{{\bf 3},3}^2 + \gamma_{{\bf 3},12}^2 
+ \left(\gamma_{{\bf 3},3}^2-\gamma_{{\bf 3},12}^2\right)\cos 2 \zeta  \right)\nonumber\\
&\ +h^2 \left(2 \gamma_1^2 - (\gamma_{{\bf 3},3}^2+\gamma_{{\bf 3},12}^2) 
- \left(\gamma_{{\bf 3},3}^2-\gamma_{{\bf 3},12}^2\right)\cos 2 \zeta \right)-4 \gamma_1^2\bigg]f = 0,
\label{eq:f2}
\end{align}
\begin{align}
&h'' + \frac{h'}{\rho} - \frac{(w-1)^2}{\rho^2}h 
-\frac{1}{8}\bigg[f^2 \left(2 \gamma_1^2 - ( \gamma_{{\bf 3},3}^2 + \gamma_{{\bf 3},12}^2) 
- \left(\gamma_{{\bf 3},3}^2-\gamma_{{\bf 3},12}^2\right)\cos 2 \zeta  \right)\nonumber\\
&\ +h^2 \left(2 \gamma_1^2 + \gamma_{{\bf 3},3}^2+\gamma_{{\bf 3},12}^2 
+ \left(\gamma_{{\bf 3},3}^2-\gamma_{{\bf 3},12}^2\right)\cos 2 \zeta \right)-4 \gamma_1^2\bigg]h = 0,
\label{eq:h2}
\end{align}
\begin{align}
&w''-\frac{w'}{\rho} - \frac{w+1}{2}f^2-\frac{w-1}{2}h^2 = 0.
\label{eq:w2}
\end{align}
Here, $\rho = m_W r /\sqrt2$, $\gamma_1 = \sqrt2 m_1/m_W$, $\gamma_{{\bf 3},3} = \sqrt2 m_{{\bf 3},3}/m_W$, and
$\gamma_{{\bf 3},12} = \sqrt2 m_{{\bf 3},12}/m_W$. 
When $m_{{\bf 3},3} = m_{{\bf 3},12}$ ($\lambda_4=0$) holds, 
the $\zeta$ dependence disappears, and these reduce to Eqs.~(\ref{eq:f1}), (\ref{eq:h1}), and (\ref{eq:w1}) with $b=0$. 
The $\zeta$ dependence
appears only in the equations for $f$ and $h$, reflecting the fact that $\lambda_4$ term is the only term which breaks the $SU(2)_{C}$ symmetry.
\begin{figure}[t]
\begin{center}
\includegraphics[width=15cm]{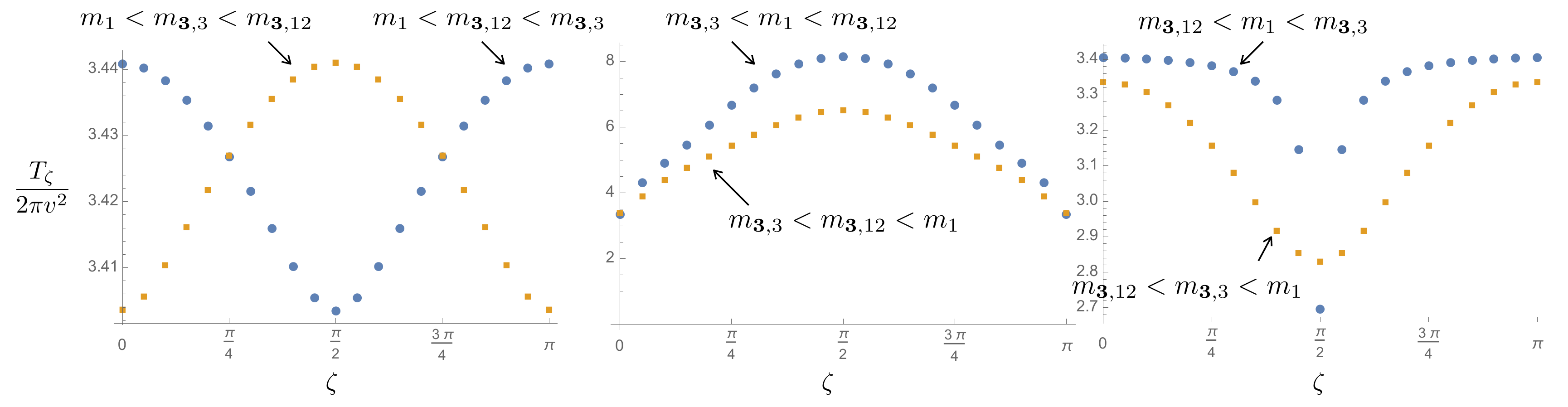}
\caption{The $\zeta$ dependence of the string tension in the model with modified potential in Eq.~(\protect\ref{eq:modified_V1}).
The left panel shows the cases with $m_1$ being the smallest while the middle and right show the cases with $m_{{\bf 3},3}$ and $m_{{\bf 3},12}$
are the smallest, respectively. 
For the left panel we take
$\frac{m_1}{m_W} =  \frac{125}{80}, m_{{\bf 3},3} = 20 m_1, m_{{\bf 3},12} = 10 m_1$ for the circle dots, and
$\frac{m_1}{m_W} =  \frac{125}{80}, m_{{\bf 3},3} = 10 m_1, m_{{\bf 3},12} = 20 m_1$ for the square dots.
For the middle panel we take
$\frac{m_{{\bf 3},3}}{m_W} =  \frac{125}{80}, m_{1} = 10 m_{{\bf 3},3}, m_{{\bf 3},12} = 20 m_{{\bf 3},3}$ for the circle dots, and
$\frac{m_{{\bf 3},3}}{m_W} =  \frac{125}{80}, m_{1} = 20 m_{{\bf 3},3}, m_{{\bf 3},12} = 10 m_{{\bf 3},3}$ for the square dots.
For the right panel we take
$\frac{m_{1}}{m_W} =  \frac{125}{80}, m_{{\bf 3},3} = 10 m_{1}, m_{{\bf 3},12} = m_{1}/2$ for the circle dots, and
$\frac{m_{{\bf 3},3}}{m_W} =  \frac{125}{80}, m_{1} = 10 m_{{\bf 3},3}, m_{{\bf 3},12} = m_{{\bf 3},3}/2$ for the square dots.
Numerical integrations are carried out over the range of $0 \le \rho \le 40$.
}
\label{fig:tension_z_w}
\end{center}
\end{figure}
We numerically solve these equations and calculate the string tension by varying the variational parameter $\zeta \in [0,\pi]$.
Numerically obtained tensions for typical cases in the six distinct regions shown in Fig.~\ref{fig:lambda_2_4} are plotted in Fig.~\ref{fig:tension_z_w}. Interestingly, the string tension behaves in two qualitatively different ways depending on the order of $m_{{\bf 3},3}$ and $m_{{\bf 3},12}$. When $m_{{\bf 3},3} < m_{{\bf 3},12}$, the string tension takes minimal value at $\zeta = 0,\pi$ and the maximum value at $\zeta = \frac{\pi}{2}$.
On the other hand, when $m_{{\bf 3},12} < m_{{\bf 3},3}$, they become upside down, namely, the global minimum is at $\zeta = \frac{\pi}{2}$,
while the global maximum are at $\zeta = 0,\pi$.
Therefore, there are two discrete string solution for the case of $m_{{\bf 3},3} < m_{{\bf 3},12}$, while there are infinitely degenerate strings
corresponding to the equator of $S^2$ moduli space of Fig.~\ref{fig:moduli}  for the case of $m_{{\bf 3},12} < m_{{\bf 3},3}$.

\subsection{$Z$- and $W$-strings at $\sin\theta_W \neq 0$}

Now, we incorporate the both effects of $U(1)_Y$ gauging discussed in Sec.~\ref{sec:U(1)Y} 
and the $SU(2)_{C}$ breaking $\lambda_4$ term introduced in the last subsection. Let us first recall that the $U(1)_Y$ gauging always makes the $Z$-string lighter compared to the $W$-string
as shown in Fig.~\ref{fig:eff_pot_Y}.
The $\lambda_4$ term also makes $Z$-string lighter when $m_{{\bf 3},3} < m_{{\bf 3},12}$, however, it makes $W$-strings lighter when $m_{{\bf 3},12} < m_{{\bf 3},3}$
as shown in Fig.~\ref{fig:tension_z_w}. Therefore, in the case of $m_{{\bf 3},3} < m_{{\bf 3},12}$, we expect that the $Z$-string is always energetically favored. On the other hand, in the case of 
$m_{{\bf 3},12} < m_{{\bf 3},3}$, it is rather non-trivial whether the $Z$-string is lighter or heavier compared to the $W$-string since there are two competing effects coming from the $U(1)_Y$ gauge interaction and the $\lambda_4$ term. In order to examine which string, $Z$ or $W$, is energetically favored, we again make the ansatz  given in Eq.~(\ref{eq:Z_string}) for $H_0$ and those given in Eq.~(\ref{eq:ansatz_B_W}) for $W_\mu$ and $B_\mu$.
Then we plug these together with the rotated covariant derivative given in Eq.~(\ref{eq:rotcov}) into the Lagrangian, we obtain the following equations that minimize the energy:
\begin{align}
&f'' + \frac{f'}{\tilde\rho} -\frac{\left((1+w)^2 - 2 (1+w)b \cos \zeta +b^2\right)}{4 \tilde\rho^2}f \nonumber\\ 
&\ -\frac{1}{8}\bigg[f^2 \left(2 \tilde\gamma_1^2 + \tilde\gamma_{{\bf 3},3}^2 + \tilde\gamma_{{\bf 3},12}^2 
+ \left(\tilde\gamma_{{\bf 3},3}^2-\tilde\gamma_{{\bf 3},12}^2\right)\cos 2 \zeta  \right)\nonumber\\
&\ +h^2 \left(2 \tilde\gamma_1^2 - (\tilde\gamma_{{\bf 3},3}^2+\tilde\gamma_{{\bf 3},12}^2) - \left(\tilde\gamma_{{\bf 3},3}^2-\tilde\gamma_{{\bf 3},12}^2\right)
\cos 2 \zeta \right)-4 \tilde\gamma_1^2\bigg]f = 0,
\end{align}
\begin{align}
&h'' + \frac{h'}{\tilde\rho} -\frac{\left((1-w)^2 + 2 (1-w)b \cos\zeta +b^2\right)}{4 \tilde\rho^2}h \nonumber\\
&\ -\frac{1}{8}\bigg[f^2 \left(2 \tilde\gamma_1^2 - 
( \tilde\gamma_{{\bf 3},3}^2 + \tilde\gamma_{{\bf 3},12}^2) - \left(\tilde\gamma_{{\bf 3},3}^2-\tilde\gamma_{{\bf 3},12}^2\right)\cos 2 \zeta  \right)\nonumber\\
&\ +h^2 \left(2 \tilde\gamma_1^2 + \tilde\gamma_{{\bf 3},3}^2+\tilde\gamma_{{\bf 3},12}^2 
+ \left(\tilde\gamma_{{\bf 3},3}^2-\tilde\gamma_{{\bf 3},12}^2\right)\cos 2 \zeta \right)-4 \tilde\gamma_1^2\bigg]h = 0,
\end{align}
\be
w''-\frac{w'}{\tilde\rho} 
- \frac{1}{2} \cos^2\theta_W \left( (1+w - b\cos\zeta)f^2 -  (1-w + b\cos \zeta)h^2 \right) = 0,\ee
\be
b''-\frac{b'}{\tilde\rho} 
+ \frac{1}{2} \sin^2\theta_W \left( ((1+w)\cos\zeta - b)f^2 - ((1-w)\cos \zeta + b)h^2 \right) = 0.
\ee
Here, we defined $\tilde \rho = m_Z r /\sqrt2 $, $\tilde \gamma_X = \sqrt2  m_X/m_Z$.
Note that when $\zeta = \pi/2$, $b$ can be taken to be identically $0$. Then the above equations reduce to the following simplified equations for the $W$-string
\begin{align}
&f'' + \frac{f'}{\tilde\rho} -\frac{(1+w)^2}{4 \tilde\rho^2}f
-\frac{1}{4}\bigg[f^2 \left(\tilde\gamma_1^2  + \tilde\gamma_{{\bf 3},12}^2\right)
+h^2 \left( \tilde\gamma_1^2 - \tilde\gamma_{{\bf 3},12}^2 \right)-2 \tilde\gamma_1^2\bigg]f = 0,
\label{eq:W_reduce1}\\
&h'' + \frac{h'}{\tilde\rho} -\frac{(1-w)^2 }{4 \tilde\rho^2}h 
-\frac{1}{4}\bigg[f^2 \left( \tilde\gamma_1^2 - \tilde\gamma_{{\bf 3},12}^2  \right)
+ h^2 \left(\tilde\gamma_1^2 +\tilde\gamma_{{\bf 3},12}^2 \right)- 2 \tilde\gamma_1^2\bigg]h = 0,
\label{eq:W_reduce2}\\
&w''-\frac{w'}{\tilde\rho} 
- \frac{1}{2} \cos^2\theta_W \left[ (1+w )f^2 -  (1-w )h^2 \right] = 0.
\label{eq:W_reduce3}
\end{align}
Note that these are independent of $\tilde\gamma_{{\bf 3},3}$.

\begin{figure}[t]
\begin{center}
\includegraphics[width=12cm]{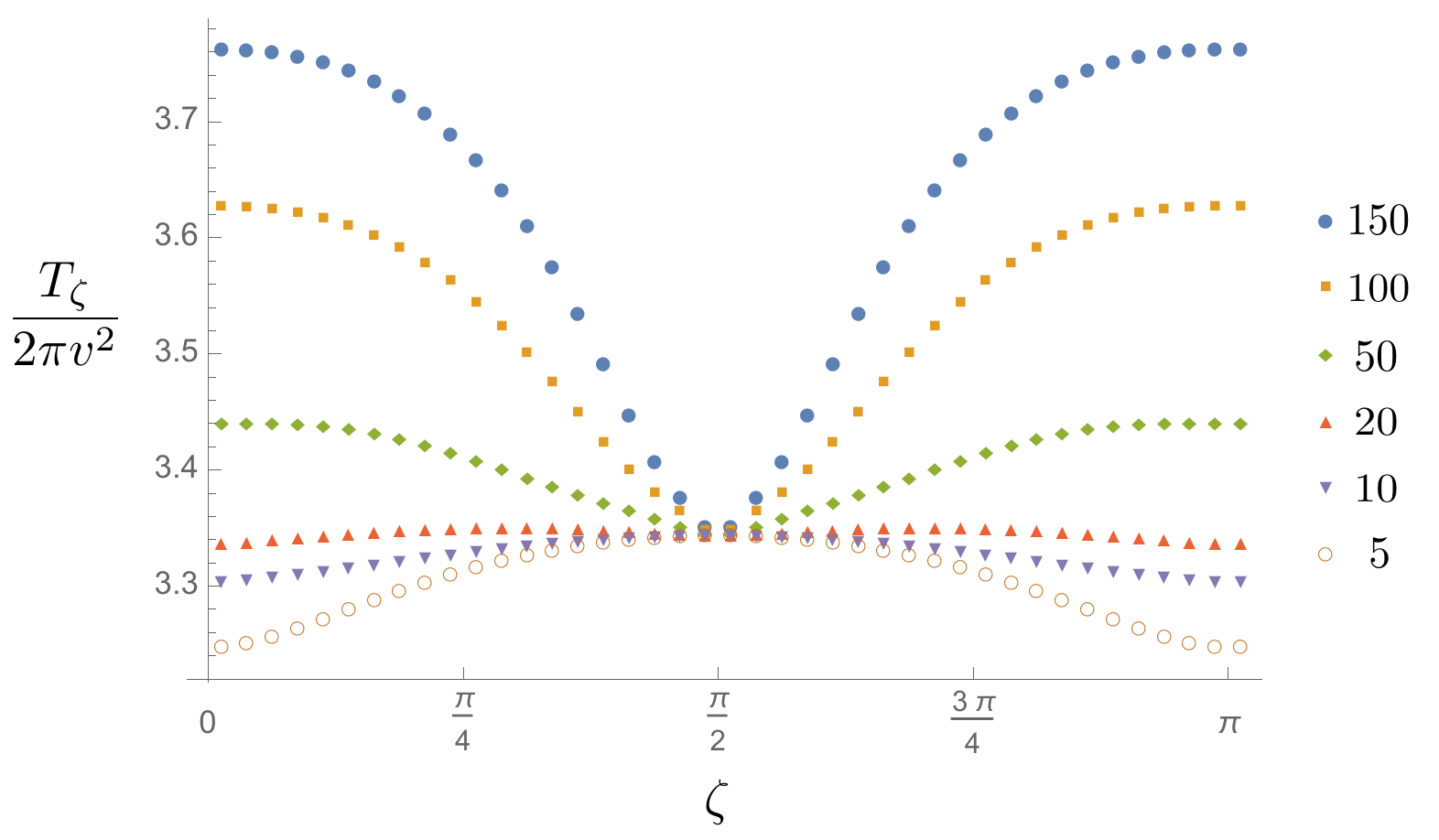}
\caption{
$\zeta$ dependence of the string tension for various values of $\tilde \gamma_{{\bf 3},3}/\tilde\gamma_1$: $\tilde \gamma_{{\bf 3},3}/\tilde\gamma_1 = 5,10,20,50,100,150$ are indicated by different symbols.
 $\tilde \gamma_1 = \frac{125}{91}$, $\tilde \gamma_{{\bf 3},12}/\tilde\gamma_1= 10$ and $\sin^2\theta_W = 0.23$ are fixed.
}
\label{fig:z_w}
\end{center}
\end{figure}

Fig.~\ref{fig:z_w} shows the $\zeta$ dependence of the string tension for various values of $\tilde \gamma_{{\bf 3},3}$ with fixing  $\tilde \gamma_1 = \frac{125}{91}$, $\tilde \gamma_{{\bf 3},12}= \frac{1250}{91}$ and $\sin^2\theta_W = 0.23$.
The $W$-string tension is common for all the choices of $\tilde\gamma_{{\bf 3},3}$ since Eqs.~(\ref{eq:W_reduce1})--(\ref{eq:W_reduce3})
are independent of $\tilde\gamma_{{\bf 3},3}$.
As is expected, the $Z$-strings are the most stable configuration for $\tilde \gamma_{{\bf 3},3} \ll  \tilde \gamma_{{\bf 3},12}$
since explicit $SU(2)_{R}$ breaking effects from both the $U(1)_Y$ gauging and the $\lambda_4$ term lower
the $Z$-string tension. On the contrary, the $W$-string is favored to the $Z$-string for  $\tilde \gamma_{{\bf 3},3} \gg  \tilde \gamma_{{\bf 3},12}$, 
where the lowering effect on the $Z$-string by $U(1)_Y$ gauging is overtaken by much stronger uplifting effect coming from the $\lambda_4$ term.
For  $\tilde \gamma_{{\bf 3},3} \simeq  \tilde \gamma_{{\bf 3},12}$, the both strings would exist as true or metastable configurations.
The result presented here, namely the possibility of energetically favored $W$-strings, has been missed since the discovery of the topologically stable $Z$-string \cite{Dvali:1993sg}.


\section{Topological $Z$-strings at $\tan\beta\neq1$}\label{sec:tanbeta}

In this section, we consider more general situation of $\tan \beta \neq 1$; 
We add the terms ${\rm Tr}[H^\dagger H \sigma_3]$ and 
${\rm Tr}[H^\dagger H \sigma_3 H^\dagger H]$ 
in addition to terms considered in Eq.~(\ref{eq:modified_V1}):
\be
V &=& -m^2 {\rm Tr}[H^\dagger H] - \mu^2 {\rm Tr}[H^\dagger H \sigma_3] 
+ \lambda_1 {\rm Tr}\!\left[(H^\dagger H)^2\right] 
+  \lambda_2 \left({\rm Tr}[H^\dagger H]\right)^2 \nonumber\\
&&
+\, \lambda_3 {\rm Tr}[H^\dagger H \sigma_3 H^\dagger H]
+ \lambda_4 {\rm Tr}[H^\dagger H \sigma_3 H^\dagger H\sigma_3].
\label{eq:modified_V2}
\ee
Qualitatively different feature of this potential compared to those considered so far is the fact that $H$ can take the VEV which is not proportional to the unit matrix:
\be
\langle H \rangle = \left(
\begin{array}{cc}
v_1 & 0\\
0 & v_2
\end{array}
\right),
\ee
with
\be
v_1 &=& \sqrt{
\frac{(\lambda_1-\lambda_3+\lambda_4)m^2-(\lambda_1+2\lambda_2-\lambda_3+\lambda_4)\mu^2}{
2(\lambda_1+\lambda_4)(\lambda_1+2\lambda_2+\lambda_4)-2\lambda_3^2}
},\\
v_2 &=& \sqrt{
\frac{(\lambda_1+\lambda_3+\lambda_4)m^2+(\lambda_1+2\lambda_2+\lambda_3+\lambda_4)\mu^2}{
2(\lambda_1+\lambda_4)(\lambda_1+2\lambda_2+\lambda_4)-2\lambda_3^2}
}.
\ee
Thus we have
\be
\tan\beta \equiv \frac{v_1}{v_2} = \sqrt{\frac{(\lambda_1-\lambda_3+\lambda_4)m^2-(\lambda_1+2\lambda_2-\lambda_3+\lambda_4)\mu^2}{
(\lambda_1+\lambda_3+\lambda_4)m^2+(\lambda_1+2\lambda_2+\lambda_3+\lambda_4)\mu^2}}.
\ee

When $v_1 \neq v_2$, the $(0,1)$ and $(1,0)$ $Z$-strings are not degenerate.
Appropriate ansatz for the $Z$-strings are given by
\be
H^{(0,1)} = \left(
\begin{array}{cc}
v_1 h(r) & 0 \\
0 & v_2 f(r)e^{i\theta}
\end{array}
\right),\quad
Z_i^{(0,1)} =  \frac{2\sin^2\beta \cos\theta_W}{g}\epsilon_{ij}\frac{x^j}{r^2}(1 - w(r)),
\label{eq:01}
\ee
and
\be
H^{(1,0)} = \left(
\begin{array}{cc}
v_1 f(r)e^{i\theta} & 0 \\
0 & v_2 h(r)
\end{array}
\right),\quad
Z_i^{(1,0)} = - \frac{2\cos^2\beta\cos\theta_W}{g}\epsilon_{ij}\frac{x^j}{r^2}(1 - w(r)),
\label{eq:10}
\ee
The factors $\sin^2\beta=\frac{v_1^2}{v_1^2 + v_2^2} $ and $\cos^2\beta=\frac{v_2^2}{v_1^2 + v_2^2}$ appearing in the gauge field ansatz are necessary to minimize the tension of the $Z$-string. As a consequence,  
the magnetic flux confined inside the string is fractionally quantized \cite{Dvali:1993sg} as
\be
\Phi_Z^{(0,1)} = 2\pi \frac{\sin^2\beta\cos\theta_W}{g},\quad
\Phi_Z^{(1,0)} = -2\pi \frac{\cos^2\beta\cos\theta_W}{g}.
\label{eq:fluxes}
\ee

\begin{figure}[t]
\begin{center}
\includegraphics[width=15cm]{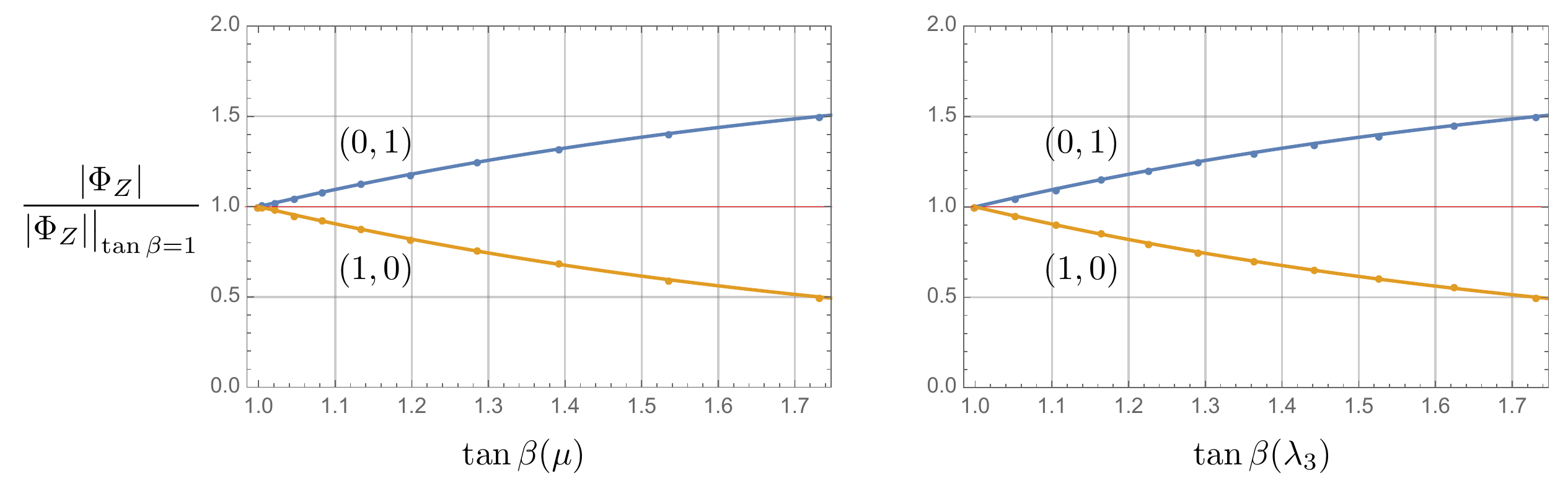}
\caption{Ratio of the $Z$ flux $|\Phi_z|$ ($\tan \beta \neq 1$) to $|\Phi_z|$ at $\tan\beta=1$. The dots are numerically obtained data, while curves are drawn by using analytical expressions in Eq.~(\protect\ref{eq:fluxes}). For numerical computation, we set the parameters $g=g'=m=\lambda_1=\lambda_2=\lambda_4=1$. For the left panel, we fixed $\lambda_3 (=0)$ and varied the value of $\mu$, while for the right panel, we fixed $\mu (=0)$ and varied the value of $\lambda_3$. In both panel, ratios are plotted as functions of $\tan \beta$.}
\label{fig:Zflux}
\end{center}
\end{figure}

We solved the equations of motion for the potential given in Eq.~(\ref{eq:modified_V2}) with the ansatz in Eqs.~(\ref{eq:01}) and (\ref{eq:10}) for various combinations of input parameters. We first calculated the $Z$ flux $\Phi_z$ from those obtained gauge configurations, and compared those with analytical formula shown in Eq.~(\ref{eq:fluxes}) in Fig.~\ref{fig:Zflux}.
In the left panel, parameters are fixed except for $\mu$, while in the right, those are fixed except for $\lambda_3$. Then in the both plots, results are plotted as functions of $\tan\beta$. In both cases, numerical data sit on the analytical curves, which confirms the validity of the numerical calculations.

\begin{figure}[t]
\begin{center}
\includegraphics[width=12cm]{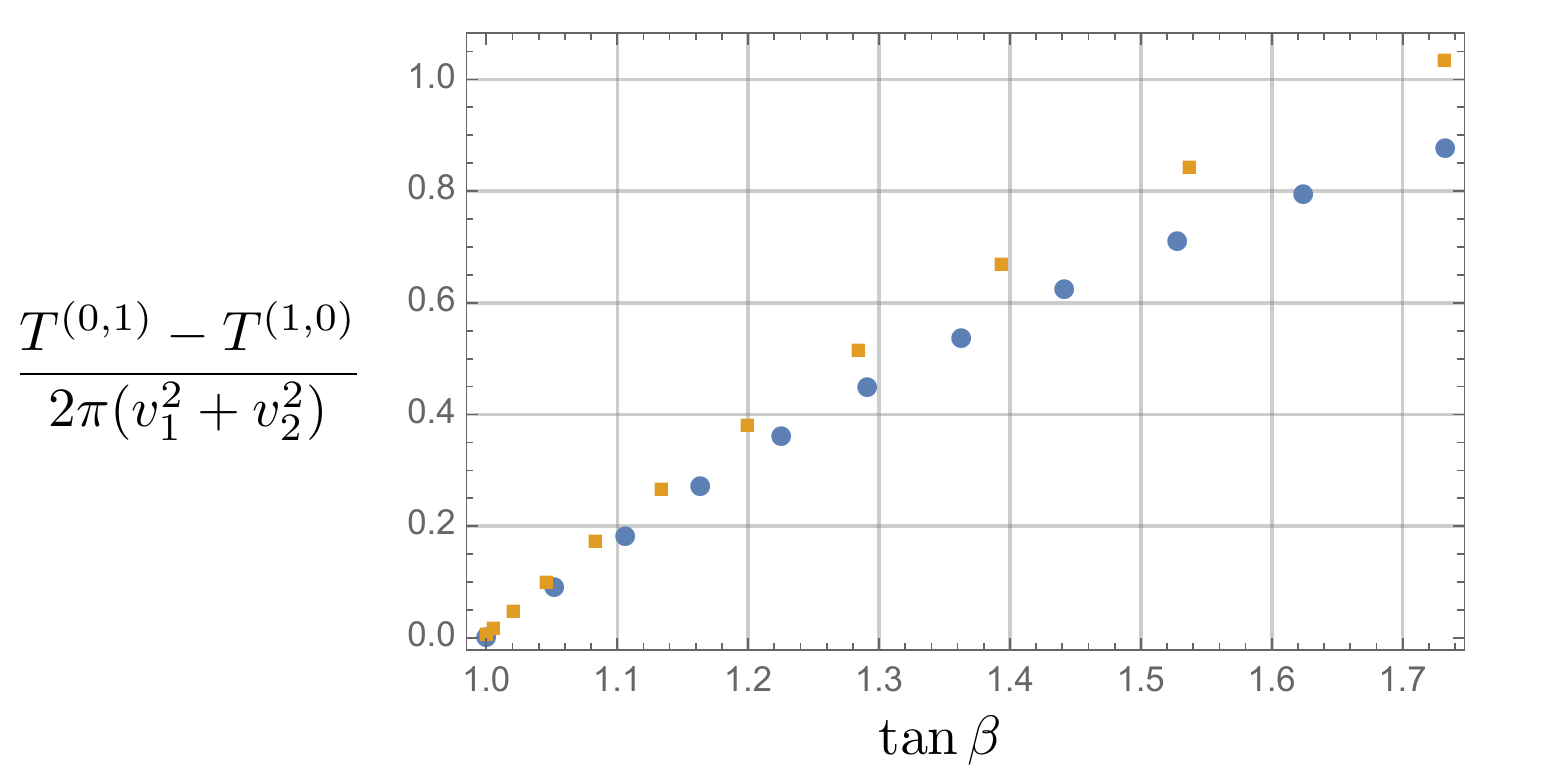}
\caption{
Difference of the tensions of $(0,1)$-string and $(1,0)$-strings normalized by $2\pi(v_1^2+v_2^2)$.
For numerical computation, we set the parameters $g=g'=\lambda_1=\lambda_2=\lambda_4=1$.
For square symbols, $\lambda_3$ is taken to be $0$ and $\mu$ is varied, while for circles, $\mu$ is taken to be $0$ and $\lambda_3$ is varied. The horizontal axis represents the value of $\tan\beta$ for given input parameters.
}
\label{fig:TZ}
\end{center}
\end{figure}

Let us next discuss the tension of the strings. 
Although the $(1,0)$- and $(0,1)$-strings have the different  $Z$ flux quanta, the dominant part of their tensions, namely
the logarithmic divergent parts are common
as
\be
\int d^2x\, {\rm Tr}[D_iH^\dagger D_i H] \simeq 2\pi \frac{v_1^2v_2^2}{v_1^2 + v_2^2} \log \Lambda.
\ee
However,  the $(1,0)$- and $(0,1)$-strings are not exactly degenerate, which can naturally be expected from the fact 
that they carry different $Z$ fluxes. Splitting in the string tensions appears in
finite subdominant part. To see this, we plot the difference of the tension of $(0,1)$ and $(1,0)$-strings (normalized by $2\pi(v_1^2+v_2^2)$) in Fig.~\ref{fig:TZ}. 
In the figure, input parameters are taken in a similar way as Fig.~{\ref{fig:Zflux}: we take $g=g'=\lambda_1=\lambda_2=\lambda_4=1$, and for square symbols, $\lambda_3$ is taken to be $0$ and $\mu$ is varied, while for circles, $\mu$ is taken to be $0$ and $\lambda_3$ is varied. Again, horizontal axis represents the value of $\tan\beta$ for given input parameters. From this figure, one can see the tension of a $(0,1)$-string is higher than that of a $(1,0)$-string for $\tan\beta > 1$. This comes from the fact that the contribution to the tension from the gauge flux in the case of a $(0,1)$-string is bigger than the case of a $(1,0)$-string, which can be understood from the magnitude of $Z$ fluxes for each case, shown in Eq.~(\ref{eq:fluxes}).

In Fig.~\ref{fig:profiles}, we show two examples of profile functions for $(0,1)$ and $(1,0)$-strings in the case of $\tan\beta= 3$.
\begin{figure}[t]
\begin{center}
\includegraphics[width=15cm]{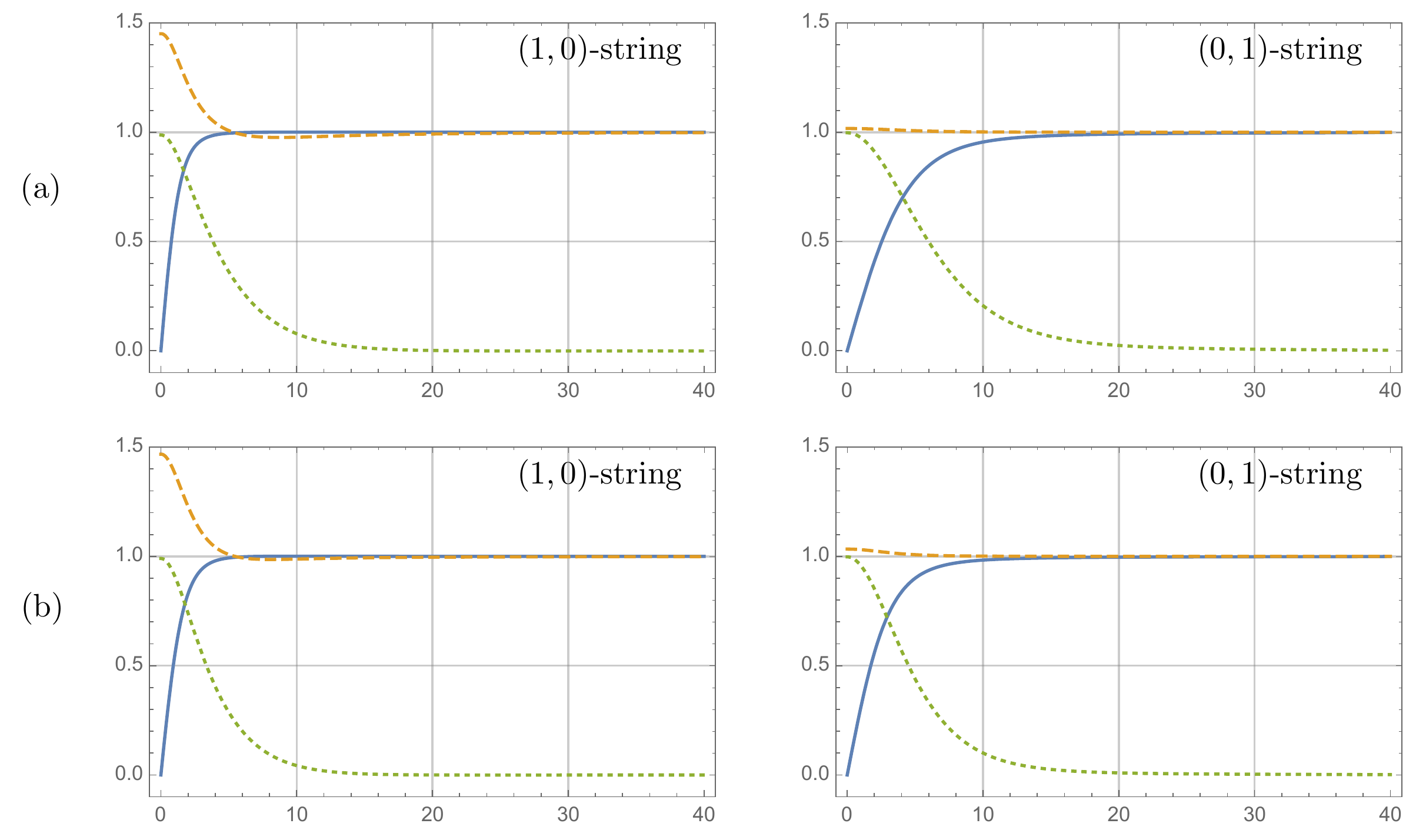}
\caption{The profile functions $f(r)$ (solid), $h(r)$ (dashed), and $w(r)$ (dotted) are shown for the $(1,0)$- and
$(0,1)$-strings. We set $\tan\beta=3$ by taking $(\mu,\lambda_3) = \left(\sqrt{\frac{2}{5}},0\right)$ for
the upper two panels (a), and $(\mu,\lambda_3) = \left(0,\frac{8}{5}\right)$ for the lower two panels (b). The other parameters
are set as $g=g'=\lambda_1=\lambda_2=\lambda_4=1$.
}
\label{fig:profiles}
\end{center}
\end{figure}

Before closing this section, it may be interesting to point out 
that fractional quantization of fluxes in Eq.~(\ref{eq:fluxes})
commonly occurs in 
the presence of multiple condensations, such as 
multi-gap superconductors 
and 
multi-component Bose-Einstein condensates.


\section{Topological $Z$-string attached by domain walls} \label{sec:string-wall}

So far, we have studied the topological $Z$- and $W$-strings in the two Higgs doublet model in the situation that the potential has a symmetry under the relative phase rotation given in Eq.~(\ref{eq:U1a}). Since the VEV of the Higgs fields spontaneously break this symmetry, the corresponding NG mode appears. Since such massless mode does not exist in nature, for phenomenologically viable model building, we need to introduce explicit $U(1)_a$ breaking terms into the Lagrangian, giving a mass to the NG mode. 
This is the CP-odd Higgs boson. 

 For the purpose of examining the effect of such $U(1)_a$ breaking terms to the $Z$-string solution studied so far, we consider the following potential:
\be
V &=& -m^2 {\rm Tr}[H^\dagger H] - \mu^2 {\rm Tr}[H^\dagger H \sigma_3] 
+ \lambda_1 {\rm Tr}\!\left[(H^\dagger H)^2\right] 
+  \lambda_2 \left({\rm Tr}[H^\dagger H]\right)^2 \nonumber\\
&&
+\, \lambda_3 {\rm Tr}[H^\dagger H \sigma_3 H^\dagger H]
+ \lambda_4 {\rm Tr}[H^\dagger H \sigma_3 H^\dagger H\sigma_3] \nonumber\\
&&- \,  m_{12}^2 \left( \det H + {\rm h.c.}\right)
+  \left(\frac{\beta_5}{2}\det H^2 + {\rm h.c.}\right).
\ee
The terms proportional to $m_{12}^2$ and $\beta_5$ are the ones that explicitly break the $U(1)_a$ symmetry.
To see how the potential depends on the relative phase of the Higgs field, let us plug $H = e^{i\alpha} {\rm diag}(v_1,\,v_2)$ into the $U(1)_a$ breaking terms:
\be
V_\xi(\alpha) &=& -2m_{12}^2 v_1v_2\cos2\alpha + \beta_5 v_1^2v_2^2\cos4\alpha \nonumber\\
&=& (v_1v_2)^2\sqrt{4\left(\frac{m_{12}^2}{v_1v_2}\right)^2 + \beta_5^2}\,
\left(-\sin\xi\cos2\alpha + \cos\xi \cos4\alpha\right),
\label{eq:Vxi}
\ee
where
\be
\sin\xi \equiv \frac{2\frac{m_{12}^2}{v_1v_2}}{\sqrt{4\left(\frac{m_{12}^2}{v_1v_2}\right)^2 + \beta_5^2}},\quad
\cos\xi \equiv \frac{\beta_5}{\sqrt{4\left(\frac{m_{12}^2}{v_1v_2}\right)^2 + \beta_5^2}},
\ee
with $\alpha \in [-\frac{\pi}{2},\frac{\pi}{2}]$ (due to the gauge equivalence\footnote{
Note that since a field with $\alpha = \delta$ (where $\delta$ is an arbitrary real value) and that with $\alpha = \delta + \pi$ are physically equivalent up to gauge transformation ($\pi$ rotation in $\sigma_3$ component of $SU(2)_W$ gauge transformation), the potential has a periodicity of $\pi$ in the direction of $\alpha$. 
} $\alpha \simeq \alpha + \pi$) and $\xi \in [0,\pi]$ (due to the our choice $m_{12}^2 \ge 0$).
This is the same form as that of the so-called double sine-Gordon potential. 

Let us now discuss the effect of the $U(1)_a$ breaking terms on the  topological $Z$-string. For this purpose, we consider the $(1,0)$-string in the following discussion. (The effect on the $(0,1)$-string can also be understood in a similar way). 
The asymptotic behavior of the Higgs field of the $(1,0)$-string configuration is given by
\be
H^{(1,0)}\big|_{r\to \infty} = 
\left(
\begin{array}{cc}
v_1 e^{i\hat \theta(\theta)} & 0 \\
0 & v_2
\end{array}
\right)
= e^{i \frac{\hat\theta(\theta)}{2}} e^{i\frac{\hat\theta(\theta)}{2}\sigma_3}
\left(
\begin{array}{cc}
v_1 & 0 \\
0 & v_2
\end{array}
\right),
\label{eq:asy10}
\ee
where $\hat\theta(\theta)$ is a function of  the angle coordinate $\theta$. In the second equality of the above, we have rewritten the phase factor of the Higgs field by the product of elements that represent the relative ($U(1)_a$) phase rotation and the common (hypercharge) phase  rotation. From this, one can see that $U(1)_a$ phase around the $(1,0)$-string varies as $ \alpha(\theta) = \frac{\hat \theta(\theta)}{2}$. Therefore, since the single valuedness of the scalar fields requires $\hat \theta(2\pi + \theta_0) - \hat \theta(\theta_0) = 2\pi$, the $U(1)_a$ phase around the $(1,0)$-string takes values from $0$ to $\pi$ (instead of $2\pi$). As we saw in previous sections, when there is no $\xi$ dependent terms in the potential, $\hat\theta(\theta)$ takes the simple form which just linearly depends on the $\theta$ as $\hat\theta(\theta) = \theta$.
However, once the potential $V_\xi(\alpha)$ is turned on, $\hat\theta(\theta)$ becomes a nontrivial function, and when $ \alpha(\theta)\, (= \frac{\hat \theta(\theta)}{2})$ passes
a potential barrier, it costs a certain additional energy. Since the region that costs additional energy density should be localized to minimize the total energy of the configuration, it forms the domain wall(s) attached the $(1,0)$-string. The number of domain walls coincides with the number of the potential barriers
which $\alpha$ passes through in the range from $\alpha = -\frac{\pi}{2}$ to $\frac{\pi}{2}$. In Fig.~\ref{fig:Vphase} (a), we plot the $\xi$ dependent part (namely the quantity in the parentheses of Eq.~(\ref{eq:Vxi})) of the potential in the range of $0 \le \xi \le \pi$ and $-\frac{\pi}{2} \le \alpha \le \frac{\pi}{2}$.
\begin{figure}[t]
\begin{center}
\includegraphics[width=15cm]{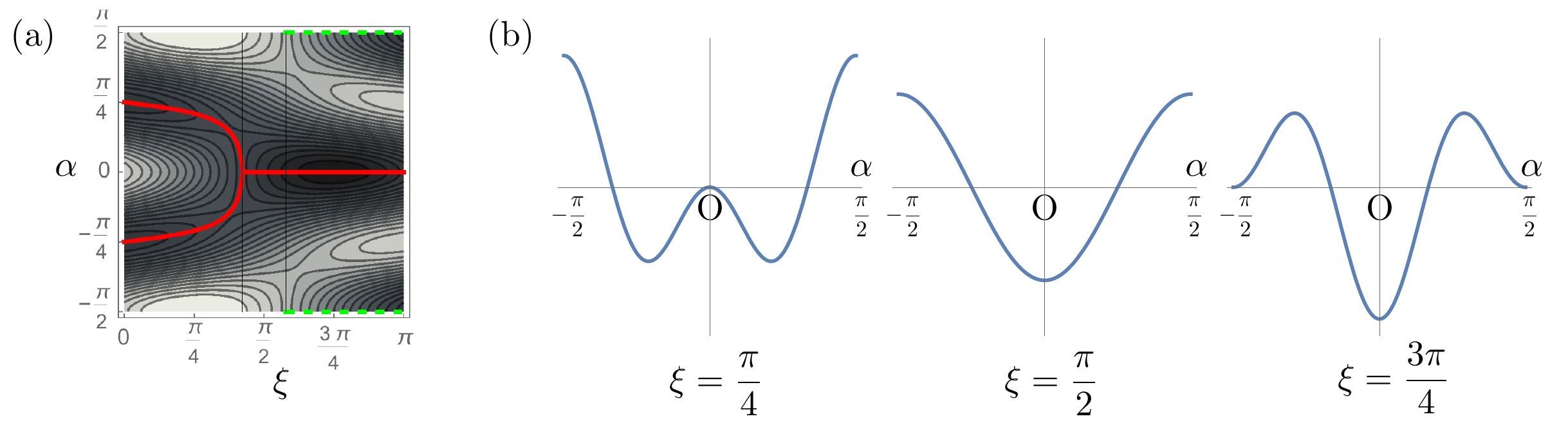}
\caption{
(a) Contour plot of the $U(1)_a$ dependent part
of the potential $V_\xi(\alpha)$. The darker the color is, the
lower the potential is. Bottom of the potential and the
local minimum are indicated by thick-dashed and
thick-solid curves, respectively. (b) Typical slices of
$V_\xi(\alpha)$ for $\alpha \in [-\frac{\pi}{2},\frac{\pi}{2}]$ with $\xi = \frac{k\pi}{4}$ ($k=1,2,3$)
are shown.
}
\label{fig:Vphase}
\end{center}
\end{figure}
The darker the color is, the lower the potential is. Bottom of the potential and the
local minimum (false vacuum) are indicated by thick-dashed and thick-solid curves, respectively. Fig.~\ref{fig:Vphase} (b) shows slices of the potential for three representative values of $\xi$, namely for $\xi = \frac{k\pi}{4}$ ($k=1,2,3$).

When the potential $V_\xi(\alpha)$ has only one global minimum, like in the case of 
$\xi=\frac{\pi}{2}$ (see the middle panel of Fig.~\ref{fig:Vphase} (b)),
$\alpha$ passes the potential barrier once when $\alpha$ moves from $-\frac{\pi}{2}$ to $\frac{\pi}{2}$. 
Therefore, in this case, one domain wall that attaches to the $(1,0)$-string appears.  Such domain wall-string 
configuration cannot be static. The string is pulled toward spacial infinity by the tension of the domain wall.
Fig.~\ref{fig:WS01} shows the (part of) field configuration (a, b), a $Z$ flux (c), and energy density (d) of a $(1,0)$-string with a domain wall at $\xi = \frac{\pi}{2}$. For concreteness, we took $\lambda_3 = \mu = 0$,
namely we have $\tan\beta=1$. The panels (a) and (b) of Fig.~\ref{fig:WS01} show square of absolute values of the upper-left (winding)
and the lower-right (unwinding) components of $H$, respectively. The winding component touches zero at the edge of the domain wall,
which evidently shows the presence of the $(1,0)$-string. The $Z$ flux is localized near the string as can be seen in the panel (c). The full energy density is plotted in the panel (d), from which one can see the energy density is concentrated near the center of the string and the domain wall attached to it.

When the potential $V_\xi(\alpha)$ has two degenerate minima, like in the case of $\xi=\pi/4$ (see the left panel of Fig.~\ref{fig:Vphase} (b)),
$\alpha$ passes the potential barrier twice when $\alpha$ moves from $-\frac{\pi}{2}$ to $\frac{\pi}{2}$. 
Therefore, in this case, two domain walls exist that attach to one $(1,0)$-string. 
Fig.~\ref{fig:WS02} shows the numerical solution of this domain wall-vortex system with $\lambda_3 = \mu = 0$ ($\tan\beta=1$). Since heights of two potential barriers are in general not equal, tensions of the domain walls are different. This can be seen from the plot of the energy density in the panel (d). This configuration is not static and the string is pulled by the heavier domain wall, unless the tensions of the two domain walls are balanced at $\xi=0$.
\begin{figure}[t]
\begin{center}
\includegraphics[width=12cm]{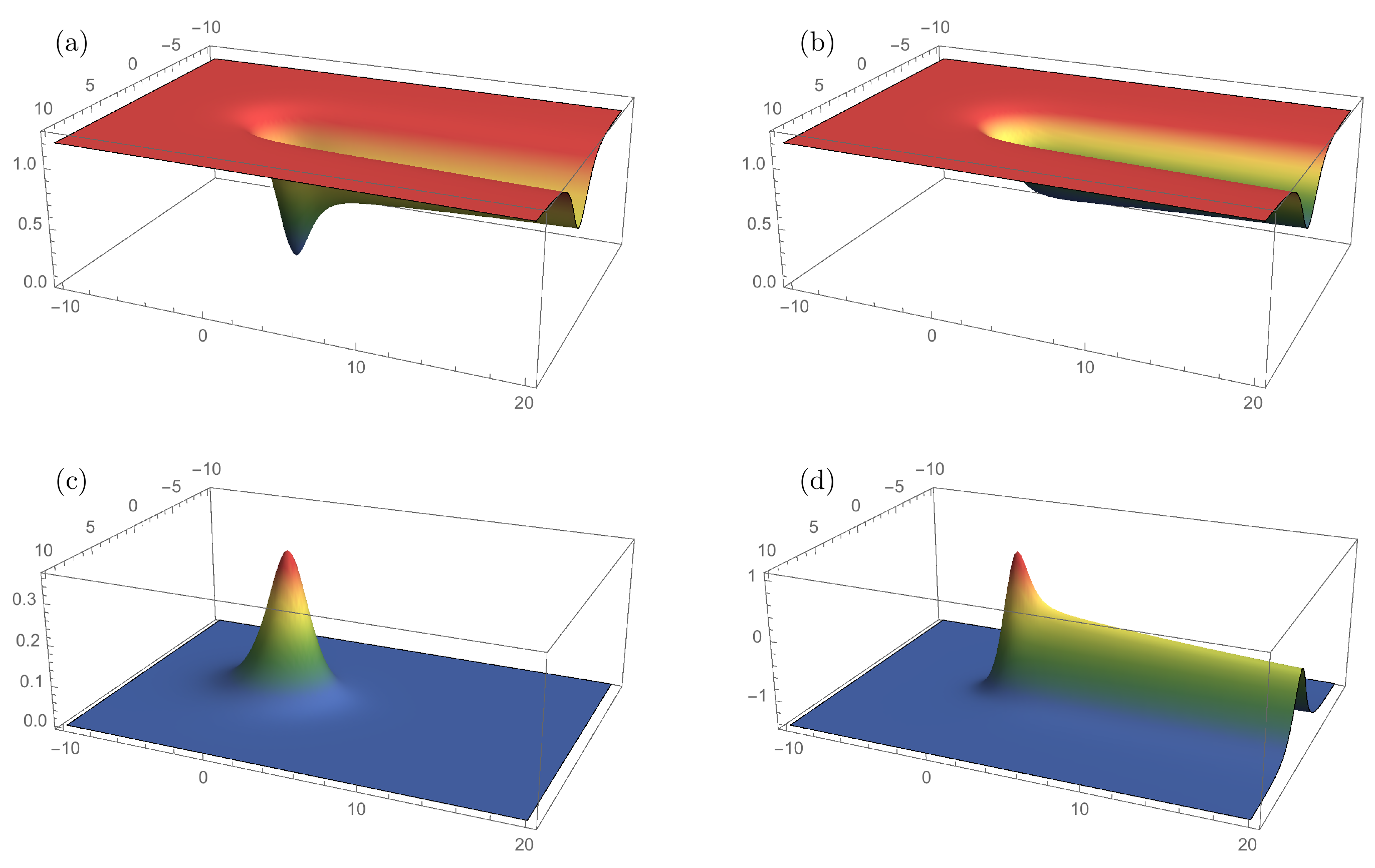}
\caption{
A snap shot of the $(1,0)$-string with a domain wall for $\xi=\frac{\pi}{2}$ on the plane orthogonal to the string. 
The panels (a) and (b) show the absolute square of the upper-left (winding) and the lower-right (unwinding) components, respectively. (c) shows the $Z$-flux and (d) shows
the full energy density.
}
\label{fig:WS01}
\end{center}
\end{figure}

\begin{figure}[t]
\begin{center}
\includegraphics[width=12cm]{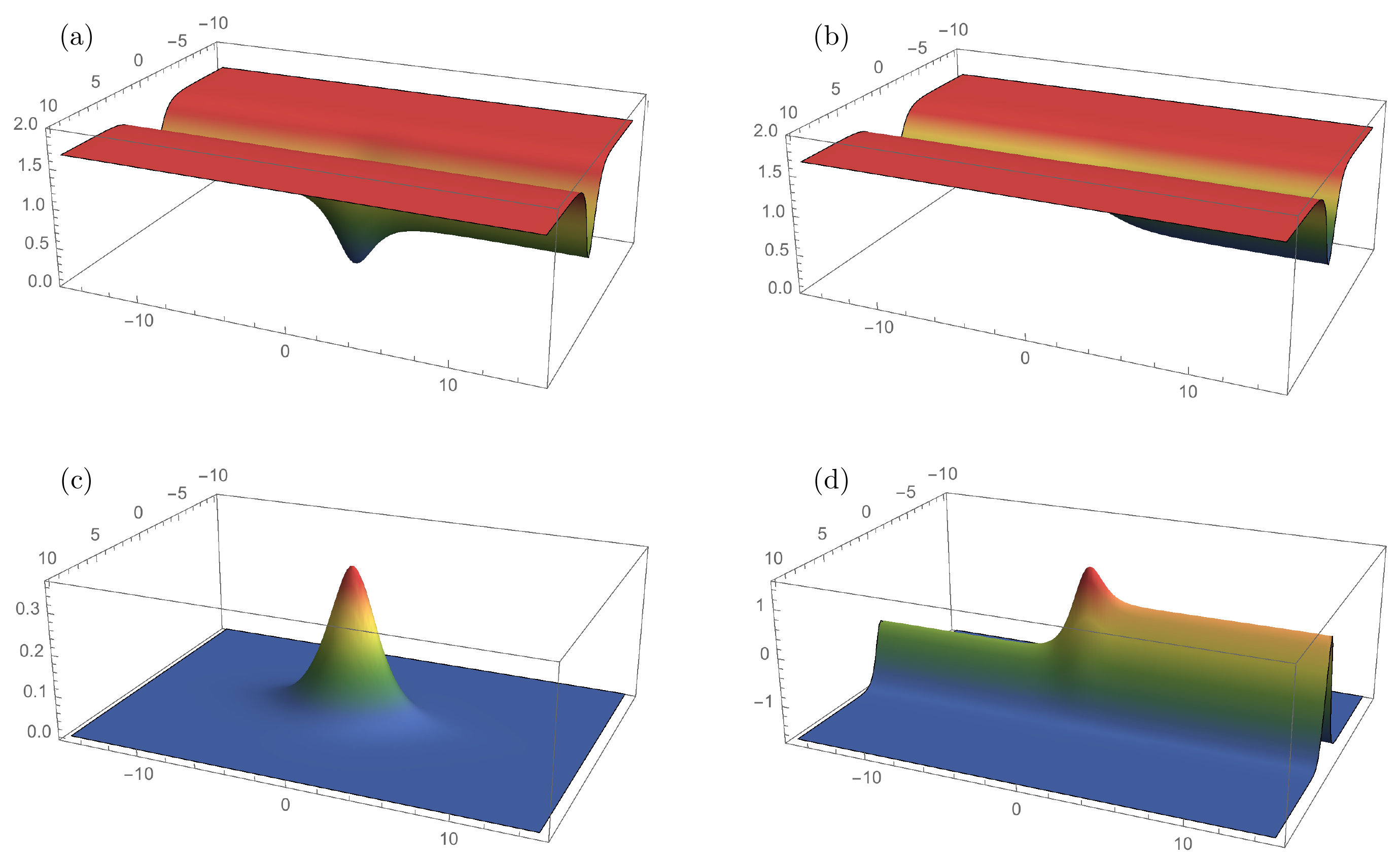}
\caption{
A snap shot of the $(1,0)$-string with two domain walls for $\xi=\pi/4$ on the plane orthogonal to the string. 
The panels (a) and (b) show the absolute square of the upper-left (winding) and the lower-right (unwinding) components, respectively. (c) shows the $Z$-flux and (d) shows
the full energy density.
}
\label{fig:WS02}
\end{center}
\end{figure}

The last example is the case when the potential $V_\xi(\alpha)$ has one global and one local minima as in the case of 
$\xi = \frac{3\pi}{4}$ (see the right panel of Fig.~\ref{fig:Vphase} (b)).  In this case, there are two potential barriers that have the same hight. Therefore, two domain walls with same tensions that attach to the $(1,0)$-string appear. However, since being in the false vacua costs additional energy, an attractive force works between the two domain walls. It is a strong confining force which is independent of
distance between the domain walls. As a consequence, two domain walls are bound together, and they end on the $(1,0)$-string from one side as shown in Fig.~\ref{fig:WS03}.
\begin{figure}[t]
\begin{center}
\includegraphics[width=12cm]{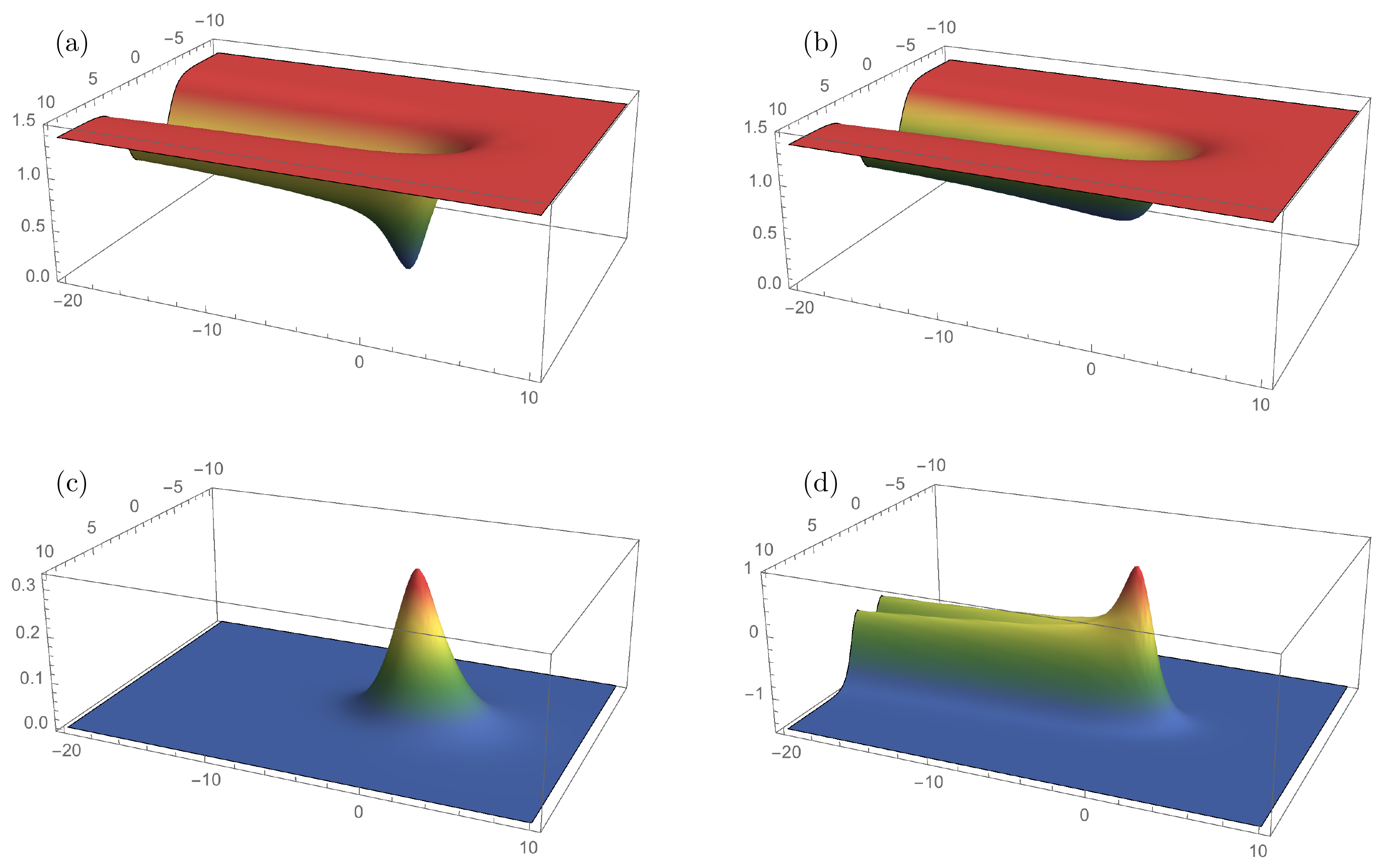}
\caption{
A snap shot of the $(1,0)$-string with a domain wall for $\xi=5\pi/4$ on the plane orthogonal to the string. 
The panels (a) and (b) show the absolute square of the upper-left (winding) and the lower-right (unwinding) components, respectively. (c) shows the $Z$-flux and (d) shows
the full energy density.
}
\label{fig:WS03}
\end{center}
\end{figure}
This type of string-domain wall system is not static since the string is pulled toward the domain walls~\cite{Eto:2018hhg}.

\section{Summary and discussion}\label{sec:summary}

In this paper, we have studied properties of the topological vortex strings and domain walls in the 2HDM in  detail. We have obtained numerical solutions of the various topological objects which were first
pointed out by Ref.~\cite{Dvali:1993sg} without concrete solutions.
Throughout the paper, we have fully utilized the two by two matrix field $H$ with which the custodial symmetry becomes transparent.
One of advantages for that is manifesting the non-Abelian moduli $S^2 \simeq SU(2)/U(1)$ 
of the topological string at the $\sin^2\theta_W = 0$ limit.
Although the non-Abelian moduli are genuine zero modes only in the special limit of the 2HDM at $\sin^2\theta_W = 0$ with the Higgs potential exactly symmetric under the custodial symmetry, the moduli space is quite useful for having a bird's-eye view of the strings in generic cases.  
For example, we have identified the $Z$-strings to the north and south poles,
and the $W$-strings to points on the equator of $S^2$. 
Gauging the $U(1)_Y$ ($\sin^2\theta_W \neq 0$) and/or modifying
the Higgs potential give rise to the potential on the moduli space. 
We have first investigated the $U(1)_Y$ gauging effect, and have found that
almost all the non-Abelian strings are lifted except for the two $Z$-strings. 
By proposing an appropriate ansatz, we have numerically derived the effective potential on the moduli space, which
correctly accounts the fact that the $Z$-strings are the most stable string solutions.
Then, we have slightly modified the Higgs potential under the restriction with $\tan\beta = 1$ being kept.
Interestingly, we have found that a certain modification of the Higgs potential results in lifting up the $Z$-strings,
which is opposite to the $U(1)_Y$ gauging effect. 
In general, these two opposite effects compete, and depending on the parameter, either the $Z$- or the $W$-string becomes the most stable. 
It is the first time to point out the possibility of the $W$-string being the most
stable string in electro-weak theories. 
We have proceeded our numerical analysis for the most generic Higgs potential for $\tan\beta \neq 1$, and obtained the numerical solutions for the most stable $Z$-strings.
At last, we have investigated effects of the additional interactions which explicitly break $U(1)_a$ which gives rise to domain walls ending on
the $Z$-strings. We have classified configurations to three cases depending on the number and type of domain walls attached to one vortex, 
and
constructed the full numerical solutions for these three string-wall composites.
It was shown that in a certain parameter region, there is a solution that the string is attached by two domain walls. 
Such system is stable and causes a cosmological problem~\cite{Eto:2018hhg}. Meanwhile, there is a parameter region where a string is attached by a domain wall from one side. Even if such system was created at a certain stage of the early Universe, it must have decayed due to the tension of the domain wall.  The remnant of this formation and decay of domain walls could be detected as a specific spectrum of the gravitational waves~\cite{Saikawa:2017hiv}, which will be studied elsewhere.

We have discussed only Higgs and gauge sectors of 2HDMs in this paper.
Including the fermion sector will be an important future step.
In general,
when fermions are coupled to Higgs fields, 
fermion zero modes are localized on 
vortices \cite{Jackiw:1981ee} and domain walls \cite{Jackiw:1975fn}. 
Fermion zero modes on conventional $Z$-strings in the SM were discussed in Refs.~\cite{Vachaspati:1992mk,Earnshaw:1994jj,Garriga:1994wb,Moreno:1994bk,Naculich:1995cb,Liu:1995at,Starkman:2000bq,Starkman:2001tc,Graham:2011fw},
and such fermions zero modes result in a lot of important phenemenological consequences.
In the case of 2HDMs, they are classified to several types (type-I, II, X, Y) depending on how fermions couple to the Higgs sector \cite{Branco:2011iw}.
It is interesting to study 
what kind of  difference exists in
fermion zero modes on a vortex and domain wall in different types of 2HDMs, 
and such a difference may result in different phenomenological consequence of 
topological solitons.

Finally, we would like to make a comment on a significant similarity underlying 
between the 2HDM and the color-flavor locked (CFL) color superconductivity
of asymptotically high density limit of QCD \cite{Alford:1998mk}. We put our emphasis on a fact that 
the two by two matrix notation $H$ again plays an
essential role for realizing this.
In the 2HDM, the $SU(2)_W$ acts on $H$ from the left, while the global $SU(2)_{R}$ symmetry acts from the right. The custodial symmetry is the vector-like $SU(2)_{C}$ symmetry combining the global
$SU(2)_W$ and $SU(2)_{R}$ as $H \to U H U^\dagger$. The $U(1)_Y$ gauge symmetry is the diagonal subgroup of 
the $SU(2)_{R}$. On the other hand, there are two order parameters in dense QCD. The one is a diquark condensate $\Phi_L$ 
of the left-handed quarks and the other is $\Phi_R$ of the right-handed quarks.
$\Phi_{L(R)}$ is a three by three matrix field which is anti-symmetric in the spin, color, flavor indices.
It transforms as $\Phi_{L(R)} \to U_C \Phi_{L(R)} U_{L(R)}$ with $U_C \in SU(3)_C$, and $U_{L(R)} \in SU(3)_{L(R)}$ 
is the chiral symmetry.
It is known that 
$\Phi_L = - \Phi_R \equiv \Phi$ holds by instanton effects \cite{Alford:1997zt,Rapp:1997zu,Rapp:1999qa}. 
Therefore, the CFL phase is effectively described
by the one condensate $\Phi$ whose transformation low is $\Phi \to U_C \Phi U_{L+R}$. 
As a consequence of the diquark condensation $\Phi = \Delta {\bf 1}_3$ in
the ground state, the QCD symmetry is spontaneously broken to the CFL global symmetry 
$\Phi \to U \Phi U^\dagger$ with $U \in SU(3)_{C+L+R}$.
Furthermore, the $U(1)_{\rm EM}$ is a diagonal subgroup of $SU(3)_{L+R}$ generated by $\lambda_8$ (the eighth component
of the Gell-Mann matrix). 
Now, the similarity is quite clear between two theories: 
$(H, SU(2)_W, SU(2)_{R}, SU(2)_{C}, U(1)_a,U(1)_Y)$ corresponds to $(\Phi, SU(3)_C, SU(3)_{L+R}, SU(3)_{C+L+R}, U(1)_B,U(1)_{\rm EM})$. In other words, we can superficially say that the 2HDM is mere two by two reduced version of the dense QCD.
It is coincidence that the topologically stable string was independently found 
in the CFL phase of dense QCD \cite{Balachandran:2005ev}, though
it was more than ten years later than the discovery of topological vortices in the 2HDM.
The fact that the string in dense QCD is the non-Abelian string was also missed at first. Later it was pointed out 
in Refs.~\cite{Nakano:2007dr,Nakano:2008dc} that it is a non-Abelian string, 
and the detail properties of the non-Abelian string in the dense QCD were
studied in Refs.~\cite{Eto:2009kg,Eto:2009bh,Eto:2009tr}.
The effect of the $U(1)$ gauging inside the flavor symmetry such as lifting up the 
moduli was found in dense QCD in Refs.~\cite{Vinci:2012mc,Cipriani:2012hr} 
(and in the Georgi-Machacek model in Ref.~\cite{Chatterjee:2018znk}). 
Thus, the topologically stable strings in the 2HDM and dense QCD are very similar 
apart from the matrix sizes; 
they are both global string carrying a non-Abelian magnetic flux 
accompanied by the non-Abelian moduli, 
${\mathbb C}P^1 (\simeq S^2)$ and ${\mathbb C}P^2$ 
in the 2HDM and dense QCD, respectively. 
 However, the potential of the 2HDM
is more generic than that of dense QCD, because 
the Ginzburg-Landau effective theory for dense QCD is more
tightly restricted by the symmetries of QCD.
The most crucial difference is the presence of $U(1)_a$ breaking terms giving rise to domain walls attached to a vortex, as discussed in detail in Sec.~\ref{sec:string-wall}. The corresponding symmetry in dense QCD is the baryon number symmetry $U(1)_{\rm B}$ which is exact, and 
there are no domain walls. 
Another important terms are
quartic interaction terms such as the term with $\lambda_4$ in the 2HDM, 
which is absent in dense QCD. 
In a certain parameter region, the $W$-string 
becomes the lightest string because of the $\lambda_4$ term as we discussed, 
but this does not happen in dense QCD.
Nevertheless, similarities between these two theories are useful since techniques in dense QCD can be imported to 2HDM. For instance, 
the low-energy effective world-sheet theory of a topological $Z$-string can be constructed 
in the same manner with dense QCD \cite{Eto:2009bh,Chatterjee:2016tml}.
A topological $Z$-string can emit and absorb $Z$ bosons, and such interaction can be obtained in the same way with that between a non-Abelian string and gluons in dense QCD \cite{Hirono:2010gq}.
Similarly, the interaction between a topological $Z$-string and the CP-odd Higgs bosons should be similar to that between a non-Abelian vortex and $U(1)_{\rm B}$ phonons in dense QCD.
A topological $Z$-string may Aharanov-Bohm (AB) scatter some particles,  
as a non-Abelian vortex AB scattering electrons and muons in dense QCD \cite{Chatterjee:2015lbf}.


\section*{Acknowledgements} 
This work is supported by the Ministry of Education, Culture, Sports, Science (MEXT)-Supported Program for the Strategic Research Foundation at Private Universities ``Topological Science'' (Grant No. S1511006).
The work is also supported in part by 
JSPS Grant-in-Aid for Scientific Research (KAKENHI Grant No. 16H03984 (M.E. and M.N.), No. 18K03655 (M.K.), No.~18H01217 (M.N.)), and also by MEXT KAKENHI Grant-in-Aid for Scientific Research on Innovative Areas ``Unification and Development of the Neutrino Science Frontier" No.~25105011 (M.K.),  ``Topological Materials Science'' No.~15H05855 (M.N.) and 
``Discrete Geometric Analysis for Materials Design" No.~JP17H06462 (M.E.) from the MEXT of Japan.

\bibliographystyle{jhep}

\end{document}